\documentclass[12pt,preprint]{aastex}

\newcommand{\degg}{\hbox{$^\circ$}}

\newcommand{\suzaku}{{\it Suzaku}}

\newcommand{\sax}{{\it Beppo-SAX}}

\newcommand{\Msun}{\hbox{$\rm\thinspace M_{\odot}$}}
\newcommand{\ls}
{\mathrel{\hbox{\rlap{\hbox{\lower4pt\hbox{$\sim$}}}\hbox{$<$}}}}
\newcommand{\gs}
{\mathrel{\hbox{\rlap{\hbox{\lower4pt\hbox{$\sim$}}}\hbox{$>$}}}}
\def\Msun{\hbox{$\rm ~M_{\odot}$}}

\received{}
\begin{document}
\title{A Compton-thick Wind in the High Luminosity Quasar, PDS 456.}
\shorttitle{The Compton-thick Wind from PDS 456.}
\shortauthors{Reeves et al.}
\author{J.N. Reeves\altaffilmark{1,2}, P.T. O'Brien\altaffilmark{3}, 
V. Braito\altaffilmark{2,3}, 
E. Behar\altaffilmark{4,5}, L. Miller\altaffilmark{6}, 
T.J. Turner\altaffilmark{4,7}, A.C. Fabian\altaffilmark{8}, 
S. Kaspi\altaffilmark{5}, R. Mushotzky\altaffilmark{4}, M. Ward\altaffilmark{9}}

\altaffiltext{1}{Astrophysics Group, School of Physical and Geographical Sciences, Keele 
University, Keele, Staffordshire, ST5 5BG, UK; jnr@astro.keele.ac.uk}
\altaffiltext{2}{Dept of Physics and Astronomy, Johns Hopkins 
University, N Charles Street, Baltimore, MD21218, USA}
\altaffiltext{3}{Dept of Physics and Astronomy, University of 
Leicester, University Road, Leicester LE1 7RH, UK}
\altaffiltext{4}{Astrophysics Division, 
NASA Goddard Space Flight Center, Greenbelt Road, Greenbelt, MD20771, USA}
\altaffiltext{5}{Dept of Physics, Technion, Haifa 32000, Israel}
\altaffiltext{6}{Dept of Physics, University of Oxford, Denys Wilkinson Building, Keble Road, Oxford OX1 3RH, UK}
\altaffiltext{7}{Joint Center for Astrophysics, University of Maryland 
Baltimore County, 1000 Hilltop Circle, Baltimore, MD21250, USA}
\altaffiltext{8}{Institute of Astronomy, Madingley Road, Cambridge CB3 0HA, UK}
\altaffiltext{9}{Dept of Physics, University of Durham, South Road, Durham DH1 3LE, UK}


\begin{abstract}

PDS 456 is a nearby (z=0.184), luminous 
($L_{\rm bol} \sim 10^{47}$\,erg\,s$^{-1}$) type I quasar. A deep 190\,ks Suzaku 
observation in February 2007 revealed the complex, broad band X-ray spectrum of 
PDS 456. The Suzaku spectrum 
exhibits highly statistically significant absorption features near 9 keV in the 
quasar rest--frame. We show that the 
most plausible origin of the absorption is from blue-shifted 
resonance ($1s-2p$) transitions of hydrogen-like iron (at 6.97 keV in the rest frame). 
This indicates that a highly ionized outflow may be present moving at near relativistic 
velocities ($\sim0.25c$). A possible hard X-ray excess  
is detected above 15\,keV with HXD (at 99.8\% confidence), 
which may arise from high column density gas
($N_{\rm H}>10^{24}$\,cm$^{-2}$) partially covering the 
X-ray emission, or through strong Compton reflection.
Here we propose that the iron K-shell 
absorption in PDS 456 is associated with a thick, possibly clumpy outflow, 
covering about 20\% of $4\pi$ steradian solid angle. The outflow is likely  
launched from the inner accretion disk, within 15--100 gravitational radii 
of the black hole. The kinetic power of the outflow may be 
similar to the bolometric luminosity of PDS 456.
Such a powerful wind could have a significant effect on the co-evolution of the 
host galaxy and its supermassive black hole, through feedback.

\end{abstract}


\section{Introduction}

Outflows are an important phenomenon in AGN and may play a key role in the
co-evolution of the massive black hole and the host galaxy. Black holes
grow by accretion and strong nuclear outflows can
quench this process by effectively shutting off the supply of matter. 
At high redshifts, such quasar winds could have provided crucial feedback
that controlled both the formation of stellar bulges and simultaneously 
self-regulated SMBH
growth, leading ultimately to the observed $M$--$\sigma$ relation for galaxies 
\citep{FM00, Gebhardt00}.
A wind, launched within a few 
gravitational radii of the black hole, can move into the host
galaxy, pushing material outwards until, at a critical black hole
mass, the ISM is effectively evacuated, quenching star
formation and stopping further SMBH growth.
Thus accretion disk outflows may help regulate the growth of the galactic bulge
and the central black hole in galaxies \citep{SR98, Fabian99, King03, 
DiMatteo05}.

Recently a number of high column density ($N_{\rm H}\sim 10^{23}-10^{24}$\,cm$^{-2}$), 
fast ($>0.1c$)
outflows have been claimed in several AGN \citep{Chartas02, Chartas03, 
Hasinger02, Reeves03, Pounds03, Dadina05, Gibson05, Markowitz06, Braito07, 
Papadakis07},
through detections of blue-shifted K-shell absorption lines of iron, observed 
at rest-frame energies greater than 7 keV.
These fast outflows are likely driven off the accretion disk by either 
radiation pressure \citep{Proga00, PK04, Sim08} or
magneto-rotational forces \citep{Kato04}, or both,
a few gravitational radii from the black
hole. In some cases the outflow rates derived can be huge, of several solar
masses per year, equivalent to $10^{45} - 10^{46}$\,erg\,s$^{-1}$ in
kinetic power \citep{Reeves03, Pounds03}, matching the radiative luminosity of the AGN
and can have a significant effect on the host galaxy evolution \citep{King03}. 
Such outflows are a likely consequence of near-Eddington accretion
\citep{KP03} and are key to understanding accretion at 
high Eddington rates.

Here we present a deep 2007 \suzaku\ observation of the nearby 
($z=0.184$, \citet{Torres97}) and luminous ($L_{\rm bol} \sim 10^{47}$\,erg\,s$^{-1}$, 
\citet{Simpson99, Reeves00}) type I quasar PDS\,456. 
A previous 40\,ks XMM-Newton observation 
revealed the presence of deep iron K-shell absorption above 7 keV, which may be 
attributed to a high column density outflow \citep{Reeves03}, 
while PDS 456 also exhibits blue-shifted absorption in the UV 
\citep{O'Brien05}.   
An analysis of the Suzaku XIS data reveals direct evidence for 
blue-shifted absorption lines in the iron K-shell band (see Sections 3 and 4), 
which may arise from a near relativistic outflow, launched 
from the inner quasar accretion disk. 
We also report on a tentative detection of hard X-ray emission above 15\,keV 
in the Suzaku HXD observation of PDS 456. 
Values of H$_{\rm 0}$=70\,km\,s$^{-1}$\,Mpc$^{-1}$,
and $\Omega_{\Lambda_{\rm 0}}=0.73$ are assumed throughout and errors are quoted at 90\% 
confidence ($\Delta\chi^{2}=2.7$), for 1 parameter of interest.

\section{Suzaku Observations of PDS 456}

PDS 456 was observed by \suzaku\ \citep{Mitsuda07} at the XIS nominal pointing 
position between 24 Feb to 1 Mar 2007 (sequence number 701056010), 
over a total duration of 370\,ks. 
A summary of observations is shown in Table\,1.
Data were analysed from the X-ray 
Imaging Spectrometer XIS \citep{Koyama07} and the PIN diodes of the  
Hard X-ray Detector HXD/PIN \citep{Takahashi07} and processed using
v2 of the \suzaku\ pipeline. Data were excluded within 
436 seconds of passage through the 
South Atlantic Anomaly (SAA) and within Earth elevation angles or Bright 
Earth angles of $<5^\circ$ and $<20^\circ$ respectively.

\subsection{XIS analysis}

XIS data were selected in $3 \times 3$ and $5 \times 5$ 
editmodes using grades 0,2,3,4,6, while hot and flickering pixels were removed 
using the {\sc sisclean} script. Spectra were extracted from within 
circular regions of 2.9\arcmin\ diameter, while background spectra were extracted from 
4 circles offset from the source and avoiding the chip corners
containing the calibration sources. The response matrix ({\sc rmf}) and ancillary response 
({\sc arfs}) files were created using the tasks {\sc xisrmfgen} and {\sc xissimarfgen},
respectively,  
the former accounting for the CCD charge injection and the latter 
for the hydrocarbon contamination on the optical blocking filter. 
Spectra from the two front illuminated XIS\,0 and XIS\,3 chips were combined to 
create a single source 
spectrum (hereafter XIS--FI), while data from the back illuminated XIS\,1 
chip were analysed separately. Data were included from 
0.45--10\,keV for the XIS--FI and 0.45--7\,keV for the XIS\,1 chip, the latter 
being optimized for the soft X-ray band. 
The net background subtracted 
source count rates were $0.521\pm0.002$\,cts\,s$^{-1}$ for XIS--FI and 
$0.413\pm0.002$\,cts\,s$^{-1}$ for XIS\,1, with a net exposure of 179.7\,ks. 
XIS background rates correspond to only 1.5\% and 3.3\% of the 
net source counts for XIS--FI and XIS\,1 respectively. The XIS spectra were subsequently binned
to a minimum energy width corresponding to approximately the $1\sigma$ XIS resolution 
of 60\,eV at 6\,keV, dropping to 30\,eV at lower energies. 
Channels were additionally grouped to achieve a 
minimum S/N of 5 per bin over the background and $\chi^{2}$ minimization 
was used for all subsequent spectral fitting. 

\subsection{HXD analysis}

The Suzaku HXD/PIN is a non-imaging instrument with a 34\arcmin\ square (FWHM) 
field of view. For weak hard X-ray sources (i.e. many AGN) where the source count rate 
is only a few percent of the background rate, it is crucial 
to correctly subtract the variable non X-ray background (i.e. the particle or 
detector background) to obtain an accurate estimate of the net source count rate.
The HXD instrument team provide two Non X-ray Background (NXB) events files 
for this purpose; the background--A model (also known as the ``quick'' background) and the 
background--D model (or ``tuned'' background). The systematic uncertainty of the background--D 
model is believed to be lower than for background--A, typically $\pm1.3$\% at the $1\sigma$
level for a net 20\,ks exposure \citep{Fukazawa09}.

To determine which model provides the most reliable estimate of the background, the 
background--A or D lightcurve was compared to the lightcurve obtained from the Earth 
occulted data (from Earth elevation angles ${\rm ELV}<-5$), in the 15--50\,keV 
band. The Earth occulted data gives a representation of the actual NXB rate, 
as this neither includes a contribution from the source nor from the Cosmic 
X-ray background (CXB). The Earth occulted data were corrected for deadtime, 
using the pseudo events file.
Figure 1 (top and middle panels) shows the comparison for each of the 
background models with the Earth rate. Specifically, we calculated the ratio 
(Earth $-$ bgd [A or D]) / bgd [A or D]; i.e. the difference between the Earth and background 
model rates normalized to the background model rate. The lightcurves are heavily binned 
(15 satellite orbits or 86.4\,ks) in order to increase signal to noise. For the 
background--A model, the model clearly underpredicts the Earth background rate, by 
$+8.8\pm1.0$\%. In contrast for background--D, the model slightly overpredicts the Earth 
rate, by the ratio $-2.4\pm1.0$\%. Thus subsequently we adopt the background D model as the 
likely more accurate background model, noting that for this observation 
it slightly overpredicts the Earth background rate, by 2.4\%, which we correct for 
in the subsequent analysis.

The background--D NXB events file 
was then used in conjunction with the screened source events file 
(from rev2 of the pipeline processing) to create a common good time interval.  
The files were thus used to calculate lightcurves (binned into 5 orbital bins or 
28.8\,ks) for both the source (deadtime corrected) and for the detector (NXB) background; 
note a $1\sigma$ systematic error of $\pm1.3$\% was included 
in the NXB rate. Figure 1 (lower panel) shows the 
PIN source lightcurve (after subtraction of the NXB) from 15--50 keV. 
The expected contribution from the diffuse (non-variable) 
Cosmic X-ray Background (CXB), using the 
spectral form of \citet{Gruber99}, was also calculated; in the 15--50 keV band this 
is expected to contribute 0.017 cts\,s$^{-1}$ to the total count rate (or a 15-50\,keV 
flux of $7.9\times10^{-12}$\,erg\,cm$^{-2}$\,s$^{-1}$). We include a 
$\pm10$\% uncertainty in the absolute normalization of the CXB.
Note a more detailed discussion of the CXB and possible 
contamination in the HXD field of view is given in Appendix\,A.

Thus the mean net source count rate, after subtraction of both the NXB and CXB components
(i.e. total background = NXB + CXB), is $(1.5\pm0.3)\times10^{-2}$\,counts\,s$^{-1}$, 
and is about 5\% of the total background count rate (0.33 counts\,s$^{-1}$). 
This corresponds to a source flux of $(7.3\pm2.4)\times10^{-12}$\,erg\,cm$^{-2}$\,s$^{-1}$ 
(including 1.3\% systematic uncertainty on the background) from 
15-50 keV, assuming a power-law photon index of $\Gamma=2$. 
In comparison the source flux derived from using background--A is higher, 
$(1.2\pm0.3)\times10^{-11}$\,erg\,cm$^{-2}$\,s$^{-1}$ from $15-50$\,keV; 
hereafter we adopt the more conservative flux derived from background--D.
This represents the first possible detection of PDS\,456 in the hard X-ray 
band above 15\,keV, at the significance level of $7\sigma$ (statistical) 
or $\sim3\sigma$ (statistical+systematic) above the background and is below the expected 
sensitivity level of the Swift-BAT all sky survey, of about 
$10^{-11}$\,erg\,cm$^{-2}$\,s$^{-1}$ \citep{Tueller08}. 

A net (deadtime-corrected) source spectrum was subsequently extracted in the 15-50 keV range
(subtracting both the NXB and CXB)  
which yielded a net PIN exposure of 164.8\,ks and the same net count rate as above 
from 15-50 keV. The deadtime rate for the source events was 6.7\%. 
The response file 
ae-hxd-pinxinome3-20080129.rsp provided by the instrument team was used 
in all the spectral fits below. The HXD/PIN spectrum was initially binned to 
a single bin, to represent a flux point between 15-50\,keV, as described above.
A finer binning of $2\sigma$ per bin above the background was subsequently 
used for the spectral fits in sections 4 and 5.
Note that a cross-normalization factor of 1.16 is included between the HXD/PIN 
and XIS, as calculated for observations of the Crab.\footnote[1]{See ftp://legacy.gsfc.nasa.gov/suzaku/doc/xrt/suzakumemo-2008-06.pdf}

\section{Detection of the Iron K-shell Absorber}

Initially we consider the X-ray spectrum of PDS 456 in the 0.45--10\,keV 
band using the XIS alone. Galactic absorption of $2\times10^{21}$\,cm$^{-2}$
\citep{DL90, Kalberla05} was adopted, modeled with the 
``Tuebingen--Boulder'' absorption model ({\sc tbabs} in {\sc xspec}) using 
the cross--sections and abundances of \citet{Wilms00}.  
Figure 2 (upper panel) shows the residuals to the XIS spectra to a single power-law 
of $\Gamma=2.35\pm0.01$ with Galactic absorption. Evidently this is a poor 
fit (fit statistic, $\chi^{2}/{\rm dof}=1004.4/391$); the 0.45--10\,keV spectrum
is concave, steepening towards lower energies, while there appears to be strong 
soft X-ray line emission observed near 0.75--0.80\,keV and absorption 
at higher energies above 7.5\,keV. 

We thus proceed to parameterize the XIS continuum with a broken power-law 
with $\Gamma_{\rm soft}=2.44\pm0.04$, a break at $E=2.0\pm0.2$\,keV and 
$\Gamma_{\rm hard}=2.24\pm0.02$ 
towards higher energies. Soft X-ray line emission is required, with 
one strong and broadened line at $E=0.91\pm0.01$\,keV (QSO rest--frame) 
with an equivalent width of ${\rm EW}=42\pm10$\,eV, intrinsic width 
of $\sigma=75\pm15$\,eV (FWHM 57000 km\,s$^{-1}$)
and a weak narrow line at $E=1.16\pm0.02$\,keV with ${\rm EW}=6\pm3$\,eV. Note 
a summary of the spectral fit parameters is shown in Table 2. 
The former line 
may be associated with either the Ne\,\textsc{ix} triplet (0.909--0.921\,keV), L-shell 
($3d-2p$) emission from Fe\,\textsc{xix} (at 0.92\,keV) or a blend of the two, 
the latter may be identified with Fe\,\textsc{xxiv} L-shell (1.16\,keV). 

The fit statistic is still 
poor ($\chi^{2}/{\rm dof}=529.3/382$, null hypothesis probability of $8.3\times10^{-7}$) and 
is significantly improved by the 
addition of two blue-shifted absorption lines observed at $7.67\pm0.04$\,keV and 
$8.13\pm0.07$\,keV (or at $9.09\pm0.05$\,keV and $9.64\pm0.08$\,keV in the quasar 
rest-frame). The fit is improved by $\Delta\chi^2=41.1$ upon first adding the 
lower energy line and further by an additional $\Delta\chi^2=29.0$ upon adding the 2nd 
higher energy line to the model. In total the fit improves by $\Delta\chi^2=70$
upon adding both lines to the model.

The line EWs are high at $-133\pm39$\,eV and $-129\pm48$\,eV respectively.
The intrinsic width of the absorption lines is marginally resolved 
$\sigma=105^{+55}_{-65}$\,eV, after subtracting the instrumental resolution of 
$\sigma=60$\,eV at Fe K. Thus the FWHM velocity is loosely constrained to be 
$8000^{+4000}_{-5000}$\,km\,s$^{-1}$. 
Note the width of the lines are tied to have the same value as each other.

The possibility that the absorption arises from a single, but 
broader line was also investigated. We note that although this cannot be firmly ruled out, 
the resulting fit is slightly worse ($\Delta\chi^{2}=12$ for 1 parameter) than for the fit with 
2 absorption lines, while the line width is substantially broadened 
($\sigma=0.40\pm0.12$\,keV or a FWHM of $\sim 30000$\,km\,s$^{-1}$).  

There is also evidence for broadened ionized iron K$\alpha$ emission 
at $7.2\pm0.2$\,keV in the quasar rest-frame 
(${\rm EW}=132\pm52$\,eV, $\sigma=0.35^{+0.65}_{-0.14}$\,keV or 
$\sim30000$\,km\,s$^{-1}$ (FWHM), $\Delta\chi^{2}=27.7$). The final fit statistic is then 
$\chi^{2}/{\rm dof}=431.5/374$.
We will return to the origin of the iron K and L-shell emission in Section 5.

We note that the absorption line 
detections are robust against the detector background of the XIS--FI CCDs, which is 
$<2$\% of the net source count rate, while the strongest emission line in the background spectrum
(Ni\,\textsc{i} K$\alpha$ at 7.47\,keV) does not coincide with 
either of the absorption lines (note that unlike in XMM-Newton EPIC-pn spectra, 
no Cu K$\alpha$ line is present in Suzaku 
XIS background spectra at 8\,keV). 
The comparison between the net source 
spectrum and background spectrum is plotted in Figure 3. 

\subsection{Statistical Significance of the Absorption Lines}

The significance of the iron K absorption lines was estimated
using multi--trial Monte Carlo simulations. The method used was identical to that  
described in \citet{Porquet04} and \citet{Markowitz06}. Specifically we generated 
3000 random XIS-FI spectra with an initial input continuum, exposure and background 
as per the actual PDS 456 data, but under the null hypothesis 
of no absorption lines. The spectra were stepped in energy
increments of 30\,eV over the 0.6--9\,keV 
energy band, i.e. where any absorption lines may be 
reasonably expected to be found, in order to map the distribution of 
$\Delta \chi^{2}$ for the trials. We ignore data below 0.6 keV as this may be effected 
by the calibration around the detector O edge and above 9 keV (observed) 
where the XIS effective area drops rapidly.

In the actual data, two absorption lines were 
found, yielding improvements to the fit statistic of $\Delta\chi^2=41.1$ and 29.0 
respectively. In the simulated spectra, these values correspond 
to 1 and 3 out of the 3000 spectra respectively. Thus the null hypothesis probability
that each of the individual lines is due to random statistical fluctuation is 
$ 3 \times 10^{-4}$ and $1 \times 10^{-3}$ respectively (i.e. $3.6\sigma$ 
and $3.2\sigma$). Indeed this significance is likely to be highly conservative, as the 
probablity was estimated
from a blind search over 3000 random spectra and stepping through 280 
energy increments (of 30\,eV) in each spectrum.

We further note that the 9\,keV rest-frame absorption is 
also independently detected in both XIS--FI CCDs, XIS\,0 and XIS\,3,  
although the detection is marginal in the 
back-illuminated XIS\,1 which has little sensitivity above 7\,keV. 
Figure 4 (top, middle) shows the XIS 0, 1, 3 spectra compared to the best fit broken power-law 
continuum, which shows the spectra from all 3 XIS are consistent within the statistical 
errors. Figure 4 (lower panel) also shows a comparison of the XIS 0 and 3 in 
the Fe K band, which illustrates the independent detection of the Fe K absorption 
in both the XIS FI CCD's. The absorption line parameters are consistent between XIS 0 
and 3; for XIS 0 the first line has an rest-frame energy 
$E=9.06\pm0.05$\,keV and $EW = -164\pm47$\,eV 
and the second line $E=9.47\pm0.05$\,keV and $EW = -149\pm60$\,eV, for XIS 3 these values are 
$E=9.12\pm0.09$\,keV, $EW = -113\pm52$\,eV and $E=9.62\pm0.15$\,keV, $EW = -101\pm64$\,eV 
respectively. The upper-limits obtained from XIS 1 are consistent with these values.

We also performed Monte Carlo simulations on 3000 spectra for XIS 0 and 3 
indepedently, as the line appears consistent in both detectors. 
As these are totally independent measurements, the resulting null probabilities are the product of the probabilities for XIS 0 and 3 seperately, which are then 
$1.2\times10^{-5}$ ($4.4\sigma$) for the lower energy line and 
$4.2\times10^{-3}$ ($2.9\sigma$) for the higher energy line.

The likelihood that the absorption is due to random 
chance is also further diminished when the 2001 XMM-Newton data are considered. Figure 2 
(lower panel) also plots the 2001 EPIC-pn data overlaid upon the Suzaku XIS spectrum. 
Absorption is clearly present in the XMM-Newton data (with a trough apparent between 
9--10\,keV, quasar rest frame), although the exact absorption profile 
may have varied. 
Indeed the possibility of deep iron K absorption 
in PDS 456 was first noted in \citet{Reeves00}, on the basis of low resolution 
RXTE PCA data and also with Beepo-SAX MECS (Vignali et al. 2000). 
Thus the coincidence of absorption in two independent detectors (XIS 0, 3) and telescopes 
(Suzaku, XMM-Newton) at very similar energies would appear to confirm the detection 
of the blue-shifted absorption in PDS\,456. 

\section{The Origin of the Iron band Absorption}

The possible origin of the iron K band absorption is now discussed.

\subsection{Identification of the Absorption Lines}

As shown above, the absorption lines detected in the XIS 
appear to be highly blue-shifted with respect to the iron K-shell 
band, with QSO rest--frame energies of $9.09\pm0.05$\,keV and $9.64\pm0.08$\,keV. Figure 2 
(lower panel) shows the XIS--FI residuals in the iron K band to the above continuum model in 
the quasar frame. Thus if either line is 
associated with the strong Fe\,\textsc{xxvi} Lyman-$\alpha$ transition at 6.97 keV, 
then this implies relativistic velocity shifts of $v_{\rm out1}=-0.26\pm0.01c$ 
($-78000\pm2000$\,km\,s$^{-1}$) and 
$v_{\rm out2}=-0.31\pm0.01c$ ($-93000\pm3000$\,km\,s$^{-1}$) 
respectively for each line, 
while if the absorption is associated with lower ionization $1s-2p$ 
transitions, e.g. Fe\,\textsc{xxv} at 6.70\,keV, the blue-shift required 
will be correspondingly higher. 
Moreover an identification without a large blue-shift is extremely unlikely;  
there are no strong line transitions in the 9.1--9.6\,keV range 
from abundant elements (e.g. Zn\,\textsc{xxx} $1s-2p$ at 9.3\,keV or 
Ni\,\textsc{xxviii} $1s-3p$ at 9.58\,keV would be undetectable barring 
extraordinarily strange abundances). Bound--free absorption 
by Fe\,\textsc{xxvi} at 9.27\,keV may appear to be plausible, however 
in a self-consistent photoionization model (see below), a strong ($>100$\,eV EW) 
Fe\,\textsc{xxvi} K$\alpha$ absorption line (at a rest-energy of 6.97\,keV, 5.89\.keV 
observed) would accompany a detectable (i.e. $\tau>0.1$) edge--like structure at 9.3\,keV;  
such a line is not present in the spectrum to a limit of $<20$\,eV in EW. We discuss this 
possibility further below.

Neither is it likely that the lines arise from local ($z=0$) ionized matter, for example 
associated with our 
galaxy or the local IGM; e.g. \citet{McKernan04, McKernan05}, but also see \citet{Reeves08}. 
In this scenario the line at 
7.67\,keV (observed frame at $z=0$) 
could only arise from K$\beta$ transitions of Fe\,\textsc{xxiii} or 
Fe\,\textsc{xxiv} from 7.6--7.8\,keV, while the 8.13\,keV (observed) line could be 
associated with either the K$\beta$ line from Fe\,\textsc{xxvi} (8.25\,keV) 
or the Ly-$\alpha$ line of Ni\,\textsc{xxviii} (8.10\,keV). However the
K$\beta$ lines of ionized iron (i.e. from Fe\,\textsc{xviii} and above) 
can be excluded, as the correspondingly stronger K$\alpha$ transitions 
between 6.5--7.0\,keV are not observed, while any Ni K-shell absorption 
will be extremely weak due to its low abundance. 

Furthermore any absorption from the local IGM would be of rather low turbulence velocity. 
For instance if a component of the local IGM or local hot bubble gas 
has a temperature of 
$T=10^{7}$\,K, then the thermal width of the lines would only be $\sigma=40$\,km\,s$^{-1}$ 
for iron atoms.
Indeed any absorption from this gas with such a low thermal width would produce very weak 
absorption lines, with equivalent widths of the order a few eV, undetectable 
with the current non-calorimeter detectors. Thus local ($z=0$) matter can be ruled out as a  
cause of the 9\,keV absorption. Instead the absorption must be associated with PDS\,456, 
possibly from highly blue-shifted absorption from an outflowing wind, as 
is discussed next.

\subsection{Photoionization Modeling of the Iron K Absorption}

We further model the Fe K absorption using a grid of \textsc{xstar} (v2.1ln9) photoionization
models. This uses the latest Fe K-shell treatment of \citet{Kallman04} 
and solar abundances \citep{GS98},
while an illuminating continuum from 1--1000 Rydbergs of $\Gamma=2.2$ is 
assumed. A large turbulence of $>1000$\,km\,s$^{-1}$ 
is required in order for the Fe K absorption lines
to be un-saturated and to be of high ($\sim 100$\,eV) equivalent width; 
such a velocity may be associated with an inner accretion 
disk wind. Here we adopt 
a high turbulence velocity of $\sigma_{\rm turb}=10000$\,km\,s$^{-1}$, 
consistent with the limits on the absorption line width, 
noting that we also tested the models (A--C) below using a lower turbulence 
grid of $3000$\,km\,s$^{-1}$, producing almost identical results.
However smaller turbulences $<1000$\,km\,s$^{-1}$ fail to reproduce the 
observed absorption line equivalent widths due to saturation.
The XIS--FI only is used in the spectral fits, as this is most sensitive towards the 
absorption above 7\,keV and we used data over the whole 0.5-10\,keV band.

The continuum is parameterized by a broken powerlaw absorbed by a Galactic 
line of sight column of gas, fitted over the 0.5-10\,keV band, as discussed previously.
The two soft X-ray Gaussian lines that are apparent near 1\,keV, 
are also included; hereafter we use this as the baseline continuum model for comparison.  
Although this provides a good description of the broad-band X-ray continuum in the XIS--FI,  
the data show clear residuals from 6--9\,keV in the iron K band (e.g. Figure 2). The 
fit statistic of $\chi^{2}/{\rm dof}=298.8/210$ is formally rejected with 
a null hypothesis probability of $5.4\times10^{-5}$. 
 
The required iron K absorption is first modeled with a single layer of 
photoionized gas fully covering the X-ray source, fitted via a 
grid of absorption models generated by Xstar and is hereafter referred to as Model A. 
We can express the model as:-
\begin{equation}
F(E) =  ({\rm BPL} + {\rm GA}_{em}) \times {\rm XSTAR}(\xi, N_{\rm H}, v_{\rm out}) 
\times e^{-\sigma(E) N_{\rm H, Gal}}
\end{equation}
where BPL is the broken powerlaw continuum, ${\rm GA}_{em}$ represents 
the Gaussian emission lines, 
XSTAR is the single layer of absorbing gas modeled via Xstar (with $N_{\rm H}$, 
$\xi$ and $v_{\rm out}$ as variable parameters), $e^{-\sigma(E) N_{\rm H, Gal}}$
is the neutral Galactic line of sight photoelectric absorption 
and $F(E)$ is the emergent energy spectrum. 
This produces a good fit to the data, 
($\chi^{2}/{\rm dof}=224.0/204$; null probability 0.151), a blend of the Fe\,\textsc{xxv} 
and Fe\,\textsc{xxvi} $1s-2p$ resonance lines models the absorption line profile; 
the Xstar absorber parameters are summarized in Table 3.
A large outflow velocity of $v_{\rm out}=-0.30\pm0.02c$ is required 
in order to reproduce the energy of the absorption lines (near 9\,keV in the rest frame). 
Furthermore a high column density ($N_{\rm H}=2.2^{+1.1}_{-1.0}\times10^{23}$\,cm$^{-2}$)
and ionization parameter (log\,$\xi=4.1\pm0.2$\footnote[2]{The units 
of $\xi$ are erg\,cm\,s$^{-1}$.}) are also required to produce 
strong highly ionized K-shell absorption lines from iron. 
The iron K-emission is included as a simple 
broad Gaussian, noting that we will 
consider more physical models for the iron K/L shell emission in Section 5 and 6.

However one Xstar zone with a single outflow velocity does not quite accurately 
model the absorption profile shown in Figure 5a (plotted from 3--10\,keV only for clarity), 
as the spacing between Fe\,\textsc{xxv} 
and Fe\,\textsc{xxvi} $1s-2p$ transitions (6.70\,keV and 6.97\,keV) 
is smaller than the observed 
separation of the absorption lines at 9.08\,keV and 9.62\,keV (rest-frame) respectively. Thus 
a better fit is subsequently obtained using two zones of absorbing matter, 
with different outflow velocities but assuming identical columns and ionizations. 
This model is referred to as 
model B, and is shown in Figure 5b. A column density of 
$N_{H}=(1.3\pm0.5) \times10^{24}$\,cm$^{-2}$ is required, 
while a high ionization parameter of ${\rm log} \xi=4.9^{+0.9}_{-0.5}$ implies
that iron is predominantly in a H-like state. 
The outflow velocities obtained are 
$v_{\rm out1}=-0.26\pm0.02c$ and $v_{\rm out2}=-0.31\pm0.02c$, 
while the fit statistic is good ($\chi^{2}/{\rm dof}=207.3/203$, null probability 0.4). 
Note that the presence of substantial columns of lower ionization gas (with 
$\log \xi < 3$) fully covering the AGN are excluded, 
as this would produce considerable absorption from L-shell iron, as well as K-shell 
lines from Mg, Si and S below 3\,keV, which are not observed in the Suzaku XIS spectrum. 
Low ionization gas, partially covering the emission below 10 keV during the Suzaku 
observation, can be excluded 
due to the lack of convex curvature in the XIS spectrum; however this is not the case for 
the 2001 XMM-Newton spectrum of PDS\,456 discussed in Section 5.3. 

We also allow both the iron emission and absorption to have the same self consistent 
velocity broadening, via convolution
with a simple Gaussian broadening function, hereafter referred to as model C.
This might be the case kinematically if the 
line emission is integrated over a smooth wide angle outflow or if the absorption is 
viewed over multiple lines of sight, e.g. if the absorber is extended with respect to the 
X-ray source, or if photons are scattered throughout the flow, as might be expected 
for the high columns observed here. This model can be expressed as:-
\begin{equation} 
F(E) =  [({\rm BPL} + {\rm GA}_{\rm em}) \times {\rm XSTAR}(\xi, N_{\rm H}, v_{\rm out})] 
\times {\rm GA}_{\rm conv} \times e^{-\sigma(E) N_{\rm H, Gal}}
\end{equation}
where ${\rm GA}_{\rm conv}$ is the Gaussian broadening convolution. 
For simplicity a single bulk outflow velocity is assumed for the absorption, while the iron  
emission/absorption is assumed to be intrinsically narrow (prior to broadening), 
except for the turbulence velocity noted above.  
The resulting model is shown in Figure 
5c, which has a velocity broadening of $26000\pm7000$\,km\,s$^{-1}$ (FWHM) and 
a net outflow velocity of $-0.29\pm0.02c$. While this represents 
a statistically adequate fit ($\chi^2/{\rm dof} = 227.9/205$, null probability 0.131), 
the model does not 
fit the profile of the absorption feature at 8 keV as well (c.f. model B), 
which appears narrower than the velocity broadening derived here. 
This might suggest that a 
homogeneous outflow is too simplistic and that significant clumping in the absorber 
may occur; this is discussed further in Section 6. 

Alternatively we attempt to model the iron K absorption with no net outflow velocity, 
in the quasar rest frame at $z=0.184$. 
This is only plausible for very low values of the turbulence velocity 
(e.g. $\sigma_{\rm turb}\sim 100$\,km\,s$^{-1}$ or less), so that the 9 keV rest-frame absorption  
is modeled by deep bound-free edges of Fe\,\textsc{xxv} 
and Fe\,\textsc{xxvi}, yet the equivalent widths of the corresponding $1s-2p$ absorption 
lines are small, as no absorption is observed in the spectrum at 
6.70 or 6.97\,keV (5.66, 5.89\,keV observed frame). Indeed even the 99.7\% confidence 
($3\sigma$) upper-limits to any narrow absorption lines at these energies are -32\,eV and
-26\,eV respectively (also including the broad iron K emission line as above).

Thus the high 
turbulence, high outflow velocity absorber from models A or B above is replaced by a low 
turbulence ($\sigma_{turb}=100$\,km\,s$^{-1}$) absorber, with no net outflow velocity.  
The resulting model is listed as model D in Table 3. The broad iron K emission line 
is also included in model D as per models A--C.
A high column density of 
$N_{\rm H}=1.2\pm0.3\times10^{24}$\,cm$^{-2}$ and 
a high ionization of ${\rm log}\xi = 4.0\pm0.2$ is required 
in order to produce a deep absorption edge near 9 keV. Nonetheless the fit obtained 
with the low turbulence absorber is still worse than for the high velocity absorber 
in model B ($\chi^2/{\rm dof}=233.3.0/205$, vs $\chi^2/{\rm dof}=207.3.0/203$) 
and the model only crudely fits the 
profile of the absorption, as shown in Figure 6. If higher turbulences are assumed, e.g. 
1000\,km\,s$^{-1}$, the model predicts significant 
resonance $1s-2p$ absorption between 5.6-6.0\,keV in the observed frame, 
which is not present in the data, resulting in a considerably worse fit of 
$\chi^2/{\rm dof}=267.5/205$. 

One other possibility is that some narrow iron K emission just happens to cover up these 
$z=0.184$ rest frame absorption lines that are not seen in the spectrum.
We denote this as model E in Table 3, which allows instead for two narrow (unresolved) 
emission lines at 
6.70\,keV and 6.97\,keV rest frame in the spectrum; the fit statistic is slightly improved
here ($\chi^{2}/{\rm dof}=228.4/204$), although this is still worse than model B 
($\chi^{2}/{\rm dof}=207.3/203$). The fit with model E is worse still if a 
higher turbulence grid is used (e.g. $\sigma=1000$\,km\,s$^{-1}$) as the predicted 
$1s-2p$ absorption lines are stronger, of EW 100 eV or more, while higher order (e.g. 
$1s-3p$) lines also become apparent and the 
fit statistic drastically worsens ($\chi^{2}/{\rm dof}=276.1/204$).

In conclusion we cannot completely rule out model D/E with zero outflow velocity, 
with these data, although this seems possible only for low turbulences 
(e.g. $100$\,km\,s$^{-1}$, as per models D and E). 
It is also the case that such high column, high ionization gas would still 
be accelerated under the radiation pressure of the quasar, with an outward momentum
rate of the order $L_{bol}/c$ (King \& Pounds 2003), as such gas would have $\tau\sim1$
to Thomson scattering and will likely not be static. We discuss this further in
Section 6.4.
Thus the most plausible origin of the absorption appears to be 
from highly blue-shifted iron K-shell lines of Fe\,\textsc{xxvi}
or Fe\,\textsc{xxv}
associated with a mildly relativistic outflow. Hereafter we 
adopt model B is the best parameterization of the iron K band absorption. 

\section{Modeling the Broad-Band Spectrum of PDS 456}

PDS 456 appears to be weakly detected in the hard X-ray band, 
as described in Section 2, with an integrated 15--50\,keV band flux of 
$7.3\pm2.4\times10^{-12}$\,erg\,cm$^{-2}$\,s$^{-1}$.
This flux is somewhat higher
than the flux measured by XIS ($3.8\pm0.1\times10^{-12}$\,erg\,cm$^{-2}$\,s$^{-1}$) in the 
2--10\,keV band. 
Indeed if we extrapolate the best fit XIS broken powerlaw 
continuum model in the previous section to 
higher energies, the predicted 15-50\,keV flux is only 
$2.1\pm0.2\times10^{-12}$\,erg\,cm$^{-2}$\,s$^{-1}$. This possible
factor of $\sim3$ increase in flux above 10\,keV is illustrated in Figure\,7, which 
shows the extrapolation of the continuum fitted to XIS out to higher energies. The 
absolute level of the HXD flux is dependent on the background systematics of the detector, 
as discussed in Section 2, which we account for as an additional
uncertainty on the HXD/PIN background level. After inclusion of this systematic 
error, the excess in HXD flux in Figure 7 appears significant at a moderate level of 
99.8\% confidence (corresponding to $\Delta\chi^{2}=13$), 
or about $3\sigma$. Further Suzaku HXD observations will 
hopefully confirm the validity of the hard X-ray detection of PDS 456.

\subsection{Reflection Dominated Models}

For now, we proceed by taking the HXD/PIN flux measurement at face value, 
noting that the fits below include the systematic error on the HXD/PIN 
background.
Compton down-scattering or ``reflection'' off optically--thick matter 
\citep{LW88, GF91}, 
e.g. the disk or torus, could in principle account 
for the hard X-ray excess. We model this with Compton
reflection off partially ionized material, 
using the \textsc{reflion} model \citep{Ross99, RF05}. 

The emergent reflected spectrum is calculated for an optically-thick
atmosphere (such as the surface of an accretion disk) of constant density
illuminated by radiation with a power-law spectrum, with a 
high-energy exponential cutoff with an e-folding energy fixed at 300 keV, i.e. 
varying as $e^{-E / 300 {\rm keV}}$. The
reflected spectrum is calculated over the range 1\,eV to 1\,MeV. Non-LTE
calculations provide temperature and ionization structures for the gas that
are consistent with the local radiation fields. 
In addition to fully-ionized
species, the following ions are included in the calculations: C\,\textsc{iii-vi}, 
N\,\textsc{iii-vii}, O\,\textsc{iii-viii}, Ne\,\textsc{iii-x}, Mg\,\textsc{iii-xii}, 
Si\,\textsc{iv-xiv}, S\,\textsc{iv-xvi}, and Fe\,\textsc{vi-xxvi}. 
Thus the model self consistently computes 
the iron line (K or L-shell) emission, as well as from other elements. 

The \textsc{reflion} model is also convolved with 
relativistic blurring from a disk around around a black hole 
\citep{Fabian89, Laor91}, to describe the reflected emission 
expected from the innermost accretion disk. 
A disk emissivity falling as $R^{-3}$ is assumed, 
along with an outer disk radius of $400R_{\rm g}$, 
where $R_{\rm g}$ is the gravitational radius. The model can be 
expressed as:- 
\begin{equation}
F(E) = [{\rm CPL} + ({\rm REF} \times \rm{KDBLUR})] 
\times \rm{XSTAR}(\xi, N_{\rm H}, v_{\rm out}) 
\times e^{-\sigma(E) N_{\rm H, Gal}} 
\end{equation}
where CPL is the illuminating cut-off powerlaw continuum, REF is the 
Compton reflected emission convolved with relativistic blurring (KDBLUR). 
Absorption from the putative high velocity outflow (denoted as XSTAR), is also included 
(modeled by an Xstar grid as per model B as described in Section 4.2) in order 
to account for the iron K-shell absorption at 9\,keV.

An adequate fit ($\chi^{2}/{\rm dof}=248.5/212$, null 
probability $4.3\times10^{-2}$) is obtained, 
with an inclination angle of $56\pm8\degg$, an inner radius of $<20 R_{\rm g}$,  
an ionization 
parameter of $\xi=190\pm95$ and a power-law photon index of $\Gamma=2.32\pm0.03$. 
The reflection fraction is then $R=\Omega/2\pi=1.3$, where $R=1$ corresponds to 
an infinite slab effectively subtending $2\pi$ steradian to the 
X-ray source.\footnote[3]{Note that we do not 
calculate the error on R, as in the \textsc{reflion} model 
the reflected ratio between the reflected 
emission and the power-law continuum is calculated over the range 1\,eV to 1\,MeV, 
with large uncertainties in the extrapolation.} A 
high Fe abundance of $\times 4$ Solar is found, driven by the 
strong iron L-shell emission near 1 keV, although this value is not  
well constrained. Overall the fit is good below 10 keV,  
however the model fails to reproduce all of the excess in the HXD above 10 keV, 
see Figure 8, 
although the precise HXD/PIN flux is subject to systematic uncertainty as discussed above. 
Thus a contribution from a reflection component towards the X-ray emission from PDS\,456 
may be present, but does not appear to account for all of the hard X-ray emission 
observed in the HXD.

\subsection{Absorption Models}

An alternative scenario is that the majority of the intrinsic continuum flux 
from PDS 456 is in fact absorbed below 10\,keV, therefore the heavily absorbed continuum 
only emerges above 10\,keV and accounts for the observed hard X-ray excess. 
This is similar to the scenario in Seyfert 2s, 
where a heavily absorbed high energy continuum is often seen \citep{Risaliti02}, 
except in PDS 456 the absorber must only partially cover 
the continuum X-ray emission in order for some direct flux below 
10\,keV to leak through, producing the 
observed continuum flux in the XIS. A similar, Compton-thick, but partially 
covering absorber has also recently been detected in the Suzaku HXD
observation of the luminous AGN (at $z=0.104$), 1H\,0419-577 \citep{Turner09}. 
Any partial covering absorption must be located close to the central AGN, i.e. much 
closer than any pc scale absorber, in order to partially cover the compact X-ray emission.

In order to model this absorption we include a second layer 
of photoionized gas via the same Xstar grid as described in Section 4.2 (with solar abundances 
and $\sigma_{\rm turb}=10000$\,km\,s$^{-1}$), in addition to the highly ionized 
absorption that is responsible for the iron K-shell lines at 9\,keV.
Thus this second layer of photoionized gas absorbs only a fraction $f$ of the 
intrinsic cut-off power-law continuum, while a fraction $1-f$ is not absorbed. 
Hence the model can be expressed as:-
\begin{equation}
F(E) = 
[({\rm CPL} \times f \times {\rm PC}) + ({\rm CPL} \times (1-f)) + {\rm GA}_{\rm em}]
\times {\rm XSTAR}(\xi, N_{\rm H}, v_{\rm out}) \times e^{-\sigma(E) N_{\rm H, Gal}}
\end{equation}
where CPL is the cut-off powerlaw continuum with an e-folding energy of 
300\,keV, PC is the layer of 
absorbing gas that partially covers the X-ray source with a covering fraction $f$, 
while XSTAR represents the highly ionized outflowing gas that is responsible for the
9\,keV iron K-shell absorption, which we assume to fully cover all of the X-ray emission.
In this model the iron K and L-shell emission is included as Gaussian line components 
represented by ${\rm GA}_{\rm em}$.
This model provides a statistically better fit ($\chi^2/{\rm dof}=217.9/209$, 
null probability 0.32), with a photon index of $\Gamma=2.27\pm0.03$. 
A large column density is required to model the heavily absorbed component above 10\,keV, 
with $N_{\rm H}=(3.5\pm1.5)\times10^{24}$\,cm$^{-2}$ and an  
ionization parameter of log\,$\xi=2.5\pm0.4$\,erg\,cm\,s$^{-1}$; 
this lower ionization parameter value (log\,$\xi<3.0$) is required 
for the resulting continuum to be largely opaque below 10\,keV, with most of the opacity 
arising from partially ionized iron (i.e. Fe\,\textsc{xxiv} and below). 
Likewise this sets the lower limit on the column density 
($N_{\rm H}>2\times10^{24}$\,cm$^{-2}$).
An outflow velocity for the partial covering absorption zone is not formally required, 
however it is not well constrained with a lower limit of $<0.2c$, 
less than for the highly ionized zone. 
The (line-of-sight) covering fraction of the absorber is $f_{cov}=0.72\pm0.12$,  
i.e. $\sim 30$\% is unabsorbed and this is the emission viewed below 10\,keV. 

Note that one drawback of the partially covering model discussed above, 
compared to the reflection model, is that the iron K and L-shell emission is 
not computed, instead we have simply parameterized the emission 
by Gaussians as discussed in Sections 3 and 4. However in the case of a Compton thick 
wind (i.e. with $N_{\rm H}>10^{24}$\,cm$^{-2}$), significant emission via Compton 
reflection off the wind would be expected and could account for observed 
emission lines as well as the iron K band absorption. 
We discuss such a model further in Section 6.3. 

\subsection{Comparison with Previous Observations}

Variations in the absorption towards PDS\,456 may also explain 
the drastic long term spectral variability of the AGN. 
Figure 9 shows the $\nu F_{\nu}$ 
spectral comparison between 
the Feb 2007 Suzaku observation with those by RXTE (1998/2001), ASCA (1998), 
XMM-Newton (2001) and Chandra (2003); see Table\,1 
for a list of observations.\footnote[4]{For clarity we show only the 1998 RXTE observation in 
Figure 9, the 
2001 observation caught the source at a similar flux level.}
Below 10 keV, PDS 456 
shows remarkable spectral variability. RXTE--1998 and XMM--2001 caught 
the AGN at a relatively high flux, as previously discussed. 
\citet{Reeves03} derived a covering fraction of 60\% 
for the absorber in the 2001 XMM-Newton observation, with a column density of 
$5.7^{+2.0}_{-2.5}\times10^{23}$\,cm$^{-2}$, an ionization parameter of 
log\,$\xi=2.5\pm0.3$ and an outflow velocity of $\sim50000$\,km\,s$^{-1}$, 
although the latter value was not well constrained in the shorter 40ks observation. 
The low resolution RXTE/PCA spectra, from both 1998 and 2001, 
also show a deep absorption trough at 8\,keV 
in the observed frame, with $\tau\sim1$ 
implying a column density for the absorber of $\sim10^{24}$\,cm$^{-2}$ 
\citep{Reeves00}. Interestingly the Beppo-SAX spectra of 
PDS 456 also show evidence for the highly ionized absorption between 8-9 keV 
\citep{Vignali2000}. We 
do not include the Beppo-SAX data in Figure 9, as the data are fairly noisy, but note 
that the spectra appear similar to those from XMM-Newton in 2001.

We re-analysed the XMM-Newton EPIC-pn spectrum from 2001, 
using the same Xstar grids as used in this current paper (with 10000\,km\,s$^{-1}$ 
turbulence). The current grids have the advantage in that they include up to date 
opacities from partially ionized iron \citep{Kallman04}; e.g. 
$1s-2p$ transitions from Fe\,\textsc{xviii-xxiv} and high order transitions 
from Fe\,\textsc{xvii} and lower. A model of the same form as the 
partial coverer in Section 5.2 was adopted. As per Suzaku XIS, 
a high ionization outflowing zone
can fit the observed dip at 8\,keV; with log\,$\xi=4.5\pm0.5$, $N_{\rm H}=5\pm2 \times 
10^{23}$\,cm$^{-2}$ and $v_{\rm out}\sim-0.25c$.
Interestingly partially covering absorption is also required in the XMM-Newton 
data to model the substantial continuum curvature present in the data from 1-10\,keV 
(see Figure 9, green points), however 
the column density is lower ($N_{\rm H}\sim1\times10^{23}$\,cm$^{-2}$) 
compared to the partial coverer in the Suzaku observation ($N_{\rm H}>10^{24}$\,cm$^{-2}$). 
The partially covering gas is low ionization (log\,$\xi<2.1$) and 
does not produce strong $1s-2p$ 
lines from  Fe\,\textsc{xviii-xxiv}, as the L-shell is filled, while its 
outflow velocity is not constrained by the pn data. The covering fraction of this 
absorber is measured to be 80\%, very similar to the Suzaku value.

In contrast to this, the Chandra/HETG observation of PDS 456 in 2003 (and also ASCA in 1998)
showed a very hard ($\Gamma=1.3$) low 
flux spectrum (i.e. XMM--2001, $F_{0.5-10}=1.1\times10^{-11}$\,erg\,cm$^{-2}$\,s$^{-1}$; 
Chandra--2003, $F_{0.5-10}=4.0\times10^{-12}$\,erg\,cm$^{-2}$\,s$^{-1}$). 
Changes in the covering and/or column of the (lower ionization) 
absorbing matter could possibly reproduce 
this spectral variability. For instance if the column density of the partial covering 
absorber was lower during the 2001 XMM-Newton observation, compared 
to the Suzaku observation (e.g. $N_{\rm H}\sim10^{23}$\,cm$^{-2}$ vs $>10^{24}$\,cm$^{-2}$), 
this would allow more of the direct continuum to be observed below 10 keV, 
resulting in a higher flux during the XMM-Newton observation and 
accounting for the convex spectral shape below 10 keV.
The lowest flux 
Chandra observation could also contain a greater contribution from the reflected 
emission, however any such component would have to be broadened as no narrow iron 
K-emission is observed.
The long-term spectral variability of PDS 456, e.g. in the 
context of a variable partial covering absorber, will be investigated in more detail 
in a subsequent paper \citep{Behar09}. 

\section{Discussion}

\subsection{The Intrinsic Luminosity of PDS 456}

One possible consequence of the high column density absorption towards PDS 456 
is that the intrinsic X-ray luminosity  
may be higher than inferred below 10\,keV; the observed 
luminosity from 2--10\,keV is only $4\times10^{44}$\,erg\,s$^{-1}$. 
In contrast the bolometric luminosity is substantially higher. From the 
optical spectrum of PDS 456 \citep{Simpson99} and correcting for the 
cosmology used in this paper,  
$\nu L_{\nu}(5100\AA)=2\times10^{46}$\,erg\,s$^{-1}$; thereby assuming a typical relation 
between the $5100\AA$ luminosity and the bolometric luminosity ($L_{\rm bol}$) of 
$L_{\rm bol}\sim9\times\nu L_{\nu}(5100\AA)$ \citep{Kaspi00}, an estimate of 
$L_{\rm bol}=1.8\times10^{47}$\,erg\,s$^{-1}$ for PDS\,456 is derived. As a consistency check, 
we calculated the integrated infra-red to UV luminosity ($L_{\rm IR-UV}$)
based on our previous spectra in the IR ($1-10 \micron$), 
optical ($3000-8000\AA$) and UV ($1200-3000\AA$) bands \citep{Simpson99,
O'Brien05}. The de-reddened $L_{\rm IR-UV}$ is $1.0\times10^{47}$\,erg\,s$^{-1}$, 
which defines a lower-limit on $L_{\rm bol}$, as the EUV band emission is not 
included. Thus hereafter we adopt $L_{\rm bol}=2\times10^{47}$\,erg\,s$^{-1}$ as the 
bolometric luminosity of PDS\,456. 

Hence the observed 2-10\,keV luminosity measured by the XIS is only 0.2\% 
of bolometric, meaning PDS 456 is X-ray faint compared to 
most quasars, where 3--5\% of bolometric may be expected for 
a typical AGN SED \citep{Elvis94}. This then suggests 
that part of the intrinsic X-ray emission may be hidden or absorbed, 
as discussed in the partial covering model, where 70\% of the 
emission may be absorbed below 10 keV by a near Compton-thick layer 
of gas. Correcting for this putative absorption below 10\,keV, the 2-10\,keV 
luminosity is then $1.2\pm0.3\times10^{45}$\,erg\,s$^{-1}$. Furthermore at the high 
column densities observed in PDS\,456, the observed continuum X-ray flux may also 
be suppressed via by a factor $e^{-\tau}$, where 
$\tau= N_{\rm H}\sigma_{T}$ and 
$\sigma_{T}=6.65\times10^{-25}$\,cm$^{2}$ is the Thomson cross-section. 
Thus for a column of $N_{\rm H}=2\times10^{24}$\,cm$^{-2}$, 
derived from the Xstar fit to the absorber, 
then $\tau=1.3$ and thus the X-ray luminosity corrected for scattering and absorption is 
$L_{2-10}=4\times 10^{45}$\,erg\,s$^{-1}$, about
2\% of the bolometric luminosity of PDS 456. This value is consistent with the 
typical $\sim3$\% ratio between $L_{2-10}$ and $L_{\rm bol}$ measured
for typical quasars \citep{Elvis94}. 
This makes PDS\,456 one of the most
luminous known nearby quasars in the X-ray band similar to 3C\,273, which is also consistent 
with its large bolometric luminosity, 
a factor of $\times 1.7$ higher than in 3C\,273 in the optical band
\citep{Simpson99}. 

\subsection{An Outflow Model for PDS 456}

Given the high ionization state (and possible partial covering) of the absorber in PDS\,456, 
it is more likely to reside close to the X-ray emission region and not at 
parsec scales commensurate with the molecular torus,  
as predicted by AGN Unified schemes \citep{Ant93} and as suggested for 
some Seyfert outflows \citep{Behar03, Blustin05}.

Thus it is not likely to be the same matter responsible for the X-ray absorption towards 
Seyfert 2 galaxies, as PDS 456 is a classic luminous type I quasar with an 
unobscured view of the optical/UV BLR \citep{Simpson99, O'Brien05}. 
The X-ray absorber in PDS 456 is probably too highly 
ionized to cause significant absorption of the UV emission, although 
an outflowing absorption component to the Ly $\alpha$ line profile  
appears in the HST-STIS spectrum ($v=14000-24000$\,km\,s$^{-1}$), 
as well as highly blue-shifted ($v=5000$\,km\,s$^{-1}$) C\,\textsc{iv} emission 
\citep{O'Brien05}. 
Thus one possibility is that 
the X-ray absorber is both part 
of a massive outflow, with density variations 
responsible for different gas layers. Dense ($n\sim10^{10}$\,cm$^{-3}$) 
clumps within the outflow can account for the possible 
Compton-thick absorption above 15\,keV, 
which partially covers the X-ray emission, 
while the less dense (and highly ionized) fully covering 
gas is responsible for the strongly blue-shifted Fe K absorption lines. 
Indeed the high ionization gas may well shield the lower ionization matter.

We consider the case of a homogeneous radial outflow from PDS 456, 
in the form of a spherical flow (or some fraction $b$ there of). 
For simplicity, we assume that 
the outflow velocity is approximately constant on the compact 
scales observed here, although 
initial acceleration must occur close to the launch radius, 
with deceleration occurring at larger radii. 
From simple conservation of mass, the outflow rate is:-
\begin{equation}
\dot{M}_{\rm out} = 4\pi b nR^{2} m_{p} v_{\rm out}
\end{equation}
(where $nR^{2} = L_{\rm ion}/\xi$). 
Here $b=1$ for a full covering, homogeneous 
spherical outflow. 

In PDS 456 the ionizing luminosity, defined by Xstar from $1-1000$\,Rydberg 
with an input continuum of $\Gamma=2.2$
is  $L_{\rm ion} = 3\times 10^{45}$\,erg\,s$^{-1}$, correcting for Galactic absorption.
This luminosity may indeed by higher by a factor of 3, e.g. if 70\% of the 
emission below 10\,keV is absorbed or scattered. 
However in the calculations below we adopt $L_{\rm ion} = 3\times 10^{45}$\,erg\,s$^{-1}$ 
as a robust lower limit to this luminosity.

As discussed in Section 4.2, the high ionization matter appears to be outflowing 
with $v_{\rm out}=-0.26c$, with an ionization parameter 
of log\,$\xi=4.9$; e.g. see model B, Table 3. 
Thus the outflow rate in PDS\,456 is of the order 
$\dot{M}_{\rm out} = 6\times 10^{27}b$\,g\,s$^{-1}$ or $100b \Msun$\,yr$^{-1}$. 
Even for a rather conservative value of $b\sim0.1$, then 
$\dot{M}_{\rm out} \sim 10\Msun$\,yr$^{-1}$. The outflow kinetic 
power is then simply $\dot{E}_{\rm out} \sim \dot{M}_{\rm out} v_{\rm out}^{2}/2$, 
which is then $2\times10^{47}b$\,erg\,s$^{-1}$. 
Thus the kinetic output of the wind may be an appreciable fraction of the 
bolometric luminosity for PDS 456 for a reasonable value of $b$. 
Note that the 
kinetic power is very sensitive to $v_{\rm out}$, effectively varying 
as $v_{\rm out}^{3}$, but note that 
we adopt the lowest value of the likely velocity 
range $v_{\rm out}=0.26-0.31c$ derived in Table 3.

There is no direct (e.g. reverberation) mass estimate for the 
black hole in PDS\, 456, however we can estimate its likely value 
from known scaling relations, derived from reverberation methods, between the AGN 
BLR virial radius and black hole mass \citep{Kaspi00, MJ02}. 
If we adopt the relation derived in \citet{MJ02}:-
\begin{equation}
M_{\rm BH} = 4.74 (\lambda L_{5100 \AA}/10^{44} {\rm erg\, s}^{-1})^{0.61\pm0.10} [{\rm FWHM}(H\beta)]^{2} \Msun
\end{equation}
where  $\lambda L_{5100 \AA}=2\times10^{46}$\,erg\,s$^{-1}$  
and ${\rm FWHM}(H\beta) = 3974\pm764$\,km\,s$^{-1}$ \citep{Torres97, Simpson99}, 
then the black mass is ${\rm log} M_{\rm BH}=9.3\pm0.4 \Msun$
for PDS 456. Alternatively, using the equivalent relation in \citet{Kaspi00} 
gives ${\rm log} M_{\rm BH}=9.5\pm0.3 \Msun$. 
The Eddington-limited luminosity for an accreting black hole 
of mass $M_{\rm BH}$ is $L_{\rm Edd}=4\pi GM m_{\rm p}c/\sigma_{T}$, 
while the Eddington accretion rate is simply 
$\dot{M}_{\rm Edd} = L_{\rm Edd}/(\eta c^{2})$, where $\eta$ is the efficiency 
of converting rest-mass into energy. 
For PDS\,456 if $M_{\rm BH}=2\times10^{9} \Msun$, then 
$L_{\rm Edd}=2\times 10^{47}$\,erg\,s$^{-1}$ and 
$\dot{M}_{\rm Edd}=50\Msun$\,yr$^{-1}$ for $\eta=0.06$ (the maximum efficiency 
for accretion onto a Schwarzschild black hole). Thus the mass outflow rate 
is likely to be a substantial fraction of the total accretion rate.

The escape radius of the outflow is simply 
$R_{\rm esc} \sim (2 c^2/v^2) R_{g} > 30R_{\rm g}$ 
(where $R_{\rm g}=GM_{\rm BH}/c^{2}$ is the gravitational radius), 
equivalent to $R_{\rm esc}=10^{16}$\,cm for 
$M_{\rm BH}=2\times 10^{9} \Msun$. For the case of a homogeneous radial outflow, 
the density varies as $n \propto R^{-2}$ and thus the ionization 
parameter $\xi=L/nR^{2}$ is largely independent of the outflow 
radius. Hence the column density viewing through the line of sight 
down to a characteristic wind radius $R_{\rm wind}$, for the case of a spherical or 
bi-conical flow viewed radially, is:-
\begin{equation}
N_{\rm H} = \int_{\rm R_{wind}}^{\infty} n(R) dR = 
\int_{\rm R_{wind}}^{\infty} (L_{\rm ion}/\xi R^{2}) dR = L_{\rm ion}/\xi R_{\rm wind}
\end{equation}
Therefore for PDS\,456, as $N_{\rm H}\sim10^{24}$\,cm$^{-2}$ and 
log\,$\xi = 4.9$, then $R_{\rm wind} = 3\times 10^{16} {\rm cm} \sim 100 R_{\rm g}$.  
Note $R_{\rm wind}$ is 
within the expected UV/BLR emission radius for PDS\,456;  
e.g. for ${\rm FWHM} ({\rm Ly}_{\alpha}) = 12000$\,km\,s$^{-1}$ \citep{O'Brien05},  
then $R_{\rm UV} \sim 1000 R_{\rm g}$. 
Note that as the highly ionized matter is located close to the 
black hole in PDS\,456, this likely excludes the possibility of low turbulence 
(and low outflow velocity) gas (model D, Section 4.2) contributing 
towards the iron K band absorption.

It is also possible that the wind is not homogeneous but is instead  
clumpy. Indeed X-ray variability is observed in PDS\,456 on rapid ($20-30$\,ks) timescales, 
e.g. as seen previously in the XMM-Newton, Beppo-SAX or RXTE observations 
\citep{Reeves00, Reeves02} and in the current Suzaku XIS observation, 
which implies a compactness 
on sizescales of several $R_{\rm g}$. 
Thus we consider the case where the absorption 
derives from clumps of matter of approximately 
constant density. In this scenario the 
maximum distance of the clumps ($R_{\rm max}$) is given by $\Delta R / R_{\rm max} < 1$, 
where $\Delta R$ is the clump thickness. Now $N_{\rm H} = n \Delta R$ 
and $n = L_{\rm ion}/\xi R_{\rm max}^{2}$ as before; thus we derive the condition that 
$R_{\rm max} < L_{\rm ion} / N_{\rm H} \xi$ and so $R_{\rm max}<100 R_{\rm g}$ for 
$\Delta R / R_{\rm max} < 1$. Thus for $R=30R_{\rm g}$, $n=3\times 10^{8}$\,cm$^{-3}$ and 
$\Delta R / R=0.3$. Indeed the fact that both high ionization ($\log \xi=4.9$) and 
lower ionization (partially covering) gas (with $\log \xi < 3$) is observed in the 
Suzaku and 2001 XMM-Newton spectra does suggest that the absorbing matter is inhomogeneous.
The gas may also become less ionized to larger radii, especially if the outer layers 
are shielded from the photoionizing X-ray source. Such matter could 
contribute towards the blue-shifted Lyman\,$\alpha$ absorption and C\,\textsc{iv} 
emission seen in the HST/STIS spectrum of PDS\,456 \citep{O'Brien05}, although with 
columns much smaller than those derived here.

\subsection{Outflow Emission and Energetics}

It may be plausible that the outflowing gas produces 
both the Fe L and K-shell emission. The emission could occur via 
transmission, but given the high column densities involved, more likely through 
reflection off the surface of the wind. To test this, we constructed a similar 
model to the Compton-thick absorber model in Section 5.2, with two absorbing layers; 
one representing a very high ionization outflowing zone 
and a lower ionization partial covering zone (e.g. representing denser clumps 
within the wind). However 
in addition to the absorption, the model also 
allows for some reflected emission (via the \textsc{reflion} model) 
from high ionization matter to represent possible 
scattering off the wind. The illuminating continuum is then just 
a single power-law of $\Gamma=2.25\pm0.03$. The reflection component is also 
convolved through a simple Gaussian velocity-broadening function 
(with $\Delta E/E$ constant with energy). This accounts for the 
line broadening clearly visible
in the 7 keV and 1 keV emission lines detected by XIS.
The model can be expressed as:-
\begin{equation}
F(E) = [({\rm CPL} \times f \times {\rm PC}) + ({\rm CPL} \times (1-f)) 
+ ({\rm REF} \times {\rm GA}_{\rm conv})]
\times \rm{XSTAR} \times e^{-\sigma(E) N_{\rm H, GAL}}
\end{equation}
where CPL is the cut-off powerlaw with $E_{\rm cut}=300$\,keV, PC is the partially 
covering absorber with a covering fraction $f$, 
XSTAR is the highly ionized outflowing absorber, REF is the reflected 
emission off partially ionized matter and ${\rm GA}_{\rm conv}$ 
corresponds to the Gaussian velocity convolution for the reflected emission.

This model is shown in Figure 10 and the fit parameters 
are summarised in Table\,4 (the ``outflowing wind'' model). 
A good fit is obtained ($\chi^{2}/{\rm dof}=217.3/211$) and the model well 
reproduces both the absorption in the XIS and HXD spectra as well as the 
ionized iron L and K-shell emission and the general shape of the 
continuum. The ionization derived for the reflector is high, 
with log\,$\xi=3.0\pm0.3$\,erg\,cm\,s$^{-1}$, which 
is intermediate between the high ionization outflowing absorber 
and the partial coverer.
The ionized reflector represents approximately 20\% 
of the broad-band flux of the power-law continuum and hence has $R=0.2$, 
but contributes less at energies above 10\,keV 
due to the high ionization of this component. Thus the scattered component may 
cover a solid angle corresponding to $b\sim0.2$ for significant
emission to be observed in the Suzaku spectrum.  We note that a value of 
$b=0.3$ was recently derived for the outflow in PG\,1211+143, via a P-Cygni like 
profile to the iron K absorption line, suggesting that the 
kinetic power of the outflow in that quasar is also likely to be similar to its total 
bolometric output \citep{PR09}.

Note that velocity broadening is required for the reflected emission, 
corresponding to 35000\,km\,s$^{-1}$, driven by the widths of the emission 
lines in the spectrum, while some net outflow of the reflector 
is also preferred (see Table 4).  
The velocity broadening may arise from the
integrated emission expected over a wind with appreciable solid angle. 
A future comparison with disk wind models being developed that 
fully incorporate self-consistent 
radiative transfer calculations for both the emission and the absorption  
\citep{Sim08, Schurch09} may prove to be interesting.

As a consistency check we compare the mass outflow rate calculated in 
Section 6.2 with that predicted via transfer of momentum through radiation 
pressure, assuming that the wind is radiatively driven. For such a high 
ionization outflow, the predominant form of momentum transfer to the wind will be 
through scattering \citep{KP03}, whereby:-
\begin{equation}
\dot{M}_{\rm out} v_{\rm out} = \frac{L_{\rm Edd}}c (1 - e^{-\tau})
\end{equation}
where $\tau\sim1$ as measured in the Suzaku spectrum of PDS\,456 and 
$L_{\rm Edd}=2\times10^{47}$\,erg\,s$^{-1}$ (see Section 6.1). Thus from the above  
$\dot{M}_{\rm out} v_{\rm out} = 4.5\times10^{36}$\,g\,cm\,s$^{-2}$ and 
as $v_{\rm out}=0.26c$, then the mass outflow rate is  
$\dot{M}_{\rm out}=6\times10^{26}$\,g\,s$^{-1}$ or $\sim10 \Msun$\,yr$^{-1}$. 
In comparison the mass outflow rate calculated in Section 6.2 (equation 5) is 
$\dot{M}_{\rm out}=6b\times10^{27}$\,g\,s$^{-1}$ and thus $b\sim0.1$ for PDS\,456, 
consistent with the estimate from the above model. Behar et al. (2009)
derive a global covering fraction of $b=0.2$ for the highly ionized absorber/reflector, 
by considering the reflected emission in all the observations of PDS 456 to date.  

If $b\sim0.2$, then the outflow is energetically significant, with 
a kinetic power of $\dot{E} = 4\times10^{46}$\,erg\,s$^{-1}$. Integrated over a 
quasar lifetime of $10^{8}$\,years and if we assume a conservative 
duty cycle of 10\% for the outflow to be active, then the total 
amount of energy released by the wind will be $\sim10^{61}$\,erg, 
plausibly exceeding the binding energy of a galaxy bulge of mass $10^{11}\Msun$ 
and velocity dispersion $\sigma=300$\,km\,s$^{-1}$ of $5\times10^{59}$\,erg. 
Thus such powerful winds could potentially cause significant feedback between the galactic 
bulge and black hole during the quasar phase. 

The duty cycle for the outflow is largely unknown, however 
we know that the iron K absorption was detected in most of the observations of 
PDS\,456 to date (see Figure 7). 
There are also several reported cases of 
high velocity outflows that are emerging in the literature, e.g. \citet{Cappi06, 
Reeves08} and references therein. The frequency of these winds and how the outflow 
depends on critical parameters (such as accretion rate) awaits a more systematic 
and statistically rigorous survey of such systems in the Suzaku and XMM-Newton 
archives. Future calorimeter based X-ray spectra with Astro-H and IXO, 
with resolution $E/\Delta E>1000$ at 6\,keV, will hopefully 
reveal a wealth of information on these outflows in the iron K band.

\subsection{Is a low velocity solution plausible for the Fe K absorption?}

Finally we consider whether it is plausible for a low velocity, low turbulence 
absorber to cause the iron K-shell absorption in PDS 456 and not a high velocity 
wind. The first apparent problem 
is that the high column, high ionization gas will naturally be accelerated under the radiation 
pressure from the central quasar. For the column densities here, the gas has an 
optical depth of $\tau \sim 1$ to Thomson scattering. In this scenario the 
outward rate of momentum transferred to the gas will be equivalent to 
$\dot{M}_{\rm out} v_{\rm out} \sim L_{\rm Edd}/c$, e.g. see King \& Pounds (2003), 
King (2003). For PDS\,456, $L_{\rm Edd} \sim L_{\rm bol} = 2\times10^{47}$\,erg\,s$^{-1}$ 
and hence the outward momentum rate for the wind is 
$\dot{M}_{\rm out} v_{\rm out} \sim 10^{37}$\,g\,cm\,s$^{-2}$. For low velocity gas, 
e.g. $v_{\rm out}=-100$\,km\,s$^{-1}$ as per model D or E, then the mass outflow rate will 
subsequently need to 
be huge for the above Thomson case, i.e. $\dot{M}_{\rm out} \sim 10^{30}$\,g\,s$^{-1}$ 
(or $10^{4} M_{\odot}$\,yr$^{-1}$) for the gas not to be outflowing with greater velocities.
This value is then inconsistent with the mass outflow rate of $\sim 10^{25}$\,g\,s$^{-1}$ 
calculated from equation (5), for $v_{\rm out}=-100$\,km\,s$^{-1}$ and log\,$\xi=4$. 
In comparison for an outflow velocity of $v_{\rm out}=-0.25c$ as per models A--C, 
the mass outflow rate derived from the Thomson case above is consistent with 
the value derived from equation (5), i.e. $\dot{M}_{\rm out} \sim 10^{27}$\,g\,s$^{-1}$. 
It may be possible for the gas to be shielded from the central source, so as not to be accelerated 
by the radiation pressure, however given the high degree of ionization of the X-ray absorber 
this appears unlikely.

Secondly, the radial distance from the black hole to the wind is unlikely to be 
substantially greater than $100R_{\rm g}$, as discussed in Section 6.2. At this distance,
it seems somewhat implausible that the highly ionized gas will have velocities as low as
100\,km\,s$^{-1}$, as is required for models D and E. 

Finally no narrow absorption has been observed in any of the X-ray or UV 
observations of PDS\,456 to date (e.g. Reeves et al. 2003; O'Brien et al. 2005; 
Behar et al. 2009). In the HST-STIS observation of PDS\,456 (O'Brien et al. 2005), 
the broadened Lyman$-\alpha$ absorption line has an outflow velocity of 
$-(14000-24000)$\,km\,s$^{-1}$ and a simple Gaussian fit to the absorption line profile 
reveals a width of $8000\pm1400$\,km\,s$^{-1}$ (FWHM) corresponding to 
$\sigma=3400\pm600$\,km\,s$^{-1}$. The XMM-Newton RGS spectrum from a 2001 observation 
has been re-analysed by Behar et al. (2009), who report evidence for absorption from 
Ne\,\textsc{ix} and Fe\,\textsc{xx}, with a corresponding outflow 
velocity of $-16000\pm1500$\,km\,s$^{-1}$ and a width of $\sigma=2500$\,km\,s$^{-1}$. 
These values are consistent with the HST-STIS UV spectrum, but while the gas is lower 
ionization and not likely to be responsible for the Fe\,\textsc{xxv-xxvi} K-shell 
absorption observed in Suzaku, it does suggest the absorption seen towards PDS\,456 
is likely high velocity and highly turbulent. 
Finally the 9 keV rest frame iron K-shell 
absorption in this Suzaku observation, as reported in Sections 3 and 4, also appears to be 
broadened. Thus it appears unlikely that narrow, low velocity gas contributes 
significantly to the X-ray absorption observed towards PDS \,456.

\section{Conclusions}

We have detected the signature of a possible Compton thick wind from PDS 456, 
a high luminosity nearby quasar ($z=0.184$) which accretes near to the Eddington limit. 
The Suzaku XIS spectrum shows evidence for statistically significant (at $>99.9$\% confidence) 
absorption in the Fe K band, 
which can be modeled by a pair of absorption lines (or one broad line) 
near 9\,keV in the quasar rest frame.
The large velocity shift, compared to the most likely identification of the 
absorption with Fe\,\textsc{xxvi} ($1s-2p$)
at 6.97 keV, implies an outflow velocity of 
$0.26-0.31c$. As has been discussed in detail, an identification with 
other atomic transitions is less likely and the large outflow velocity appears to be 
required. The large velocity implies that the mass outflow rate is high, 
of the order of several tens of Solar masses per year, with a corresponding kinetic 
power of up to 
$\sim10^{47}$\,erg\,s$^{-1}$, close to the Eddington-limited luminosity of PDS\,456. 
Such winds could be an important source of feedback regulating black hole and 
bulge growth 
\citep{King03, DiMatteo05} during the quasar growth phase in the early Universe.  

A tentative hard X-ray detection of PDS 456 above 15\,keV 
has also been made (at the $\sim 3\sigma$ level) in the HXD/PIN, at a low flux level of 
$7\times10^{-12}$\,erg\,cm$^{-2}$\,s$^{-1}$. The hard X-ray emission requires 
either a Compton-thick ($N_{\rm H}>10^{24}$\,cm$^{-2}$) 
partial covering absorber or strong reflection, 
or a combination of both, to account for the apparent hard X-ray excess. However at the 
current low significance level of the detection, a further deep observation of PDS\,456 
with Suzaku is needed to confirm the nature of the hard X-ray excess.

\section{Acknowledgements}

We would like to dedicate this paper to the memory of Professor Martin Turner CBE.
This research has made use of data obtained from the Suzaku satellite, a collaborative 
mission between the space agencies of Japan (JAXA) and the USA (NASA). We would also 
like to thank Alex Markowitz for kindly reprocessing the archived RXTE 
spectra of PDS\,456. E.B. was partially supported by NASA grant No. 08-ADP08-0076 
issued through the Astrophysics Data Analysis Program (ADP), while T.J.T was 
partially supported by NASA grant NNX08AJ41G. SK is supported at the Technion by the 
Kitzman Fellowship and by a grant from the Israel-Niedersachsen collaboration program.

\clearpage

\section*{Appendix A - Possibility of HXD Contamination}

It is important to discuss the possibility of contaminating X-ray emission from 
other sources within the HXD field of view. Nonetheless there are no known 
bright ($F_{2-10} > 10^{-13}$\,erg\,cm$^{-2}$\,s$^{-1}$) 
contaminating point sources from previous 
imaging observations of PDS 456 below 10 keV, e.g. 
with ASCA GIS, SAX/MECS, XMM-Newton EPIC which have similar fields 
of view to Suzaku/HXD. 
Above 10\,keV, no known hard X-ray sources (other than the expected detection of 
PDS 456 itself) are present within 
$2\degg$ of PDS\,456 in the RXTE slew survey \citep{Rev04} nor from the 
Integral/IBIS all sky survey \citep{Kriv07a}. The nearest known 
source detected in the Integral all sky survey is the LMXB, GX\,9+9, approximately 
$3\degg$ from PDS 456. 
There is a possibility of some weak Galactic diffuse 
contamination due to the position of PDS 456, at the very edge of the Galactic bulge,
$l=10.4\degg$, $b=+11.2\degg$), due to a population of accreting magnetic CVs. 
Results from the Integral \citep{Kriv07b} or RXTE/PCA \citep{Rev06} 
Galactic ridge surveys show that this contributes  
$<1\times10^{-12}$\,erg\,cm$^{-2}$\,s$^{-1}$ from 15-50 keV, 
integrated over the much smaller HXD/PIN field of view and thus is negligible.

There is also the contribution of 
Extragalactic diffuse (AGN) emission from the Cosmic hard X-ray Background (CXB), 
an estimate of which has already been included in the background spectrum for the HXD/PIN. 
The HXD detection of PDS 456 lies above the level expected from the CXB. 
The flux measured by the HXD, minus the non X-ray background (NXB) 
component (but not the CXB) is $1.52\pm0.20\times10^{-11}$\,erg\,cm$^{-2}$\,s$^{-1}$ 
(15-50 keV band).  In this paper we have adopted the expected 
CXB flux from the HEAO-1 measurement of 
\citet{Gruber99}, which is 
$7.9\times10^{-12}$\,erg\,cm$^{-2}$\,s$^{-1}$ from 15-50\,keV, thus the net 
intrinsic flux from PDS 456 is $7.3\pm2.4\times10^{-11}$\,erg\,cm$^{-2}$\,s$^{-1}$. 
The error includes the 1.3\% absolute uncertainty in the HXD/PIN background, 
and allows for 10\% uncertainty in the CXB determination. 
Note that the CXB flux measured by \citet{Gruber99} is also consistent with that 
obtained by \sax\ PDS by \citet{Frontera07}, who obtain a 90\% upper bound on the 
CXB level 
of $6.5\pm0.2\times10^{-8}$\,erg\,cm$^{-2}$\,s$^{-1}$\,sr$^{-1}$ (20-50\,keV) 
equivalent to $8.3\pm0.2\times10^{-12}$\,erg\,cm$^{-2}$\,s$^{-1}$, when 
normalized to the Suzaku HXD field of view and converted to the 15-50\,keV band.
We note that the Integral IBIS measurement of the CXB, by \citet{Chur07} via 
Earth occultation, is also within 10\% of the HEAO-1 measurement. The effect of 
any anisotropy of the CXB should be negligible, fluctuations are thought to be at 
the level of 5--10\% on scales of 
a square degree. As a final check of the CXB level, 
we also computed the HXD flux from the Suzaku observation of the Lockman Hole (LH), performed 
on 3--6 May 2007. The data were reduced as described in Section 2, resulting in 84.3\,ks of 
net HXD exposure. The HXD/PIN spectrum of PDS 456 
lies significantly above that of the LH emission, while the LH flux is  
consistent with the expected CXB level 
($0.9\pm0.2\times10^{-11}$\,erg\,cm$^{-2}$\,s$^{-1}$). Thus uncertainty in the CXB 
subtraction does not significantly effect the detection of PDS 456. 

One final possibility is that a bright, serendipitous AGN is located within the 
HXD field of view. From the the Swift--BAT hard X-ray selected AGN 
$logN-logS$ \citep{Tueller08}, we would randomly
expect $<5\times10^{-3}$ AGN of flux $\sim 1\times10^{-11}$\,erg\,cm$^{-2}$\,s$^{-1}$ 
per HXD field. The actual probability 
is likely much lower, as there are no other bright AGN candidates in
the soft X-ray ($<10$\,keV) or optical fields of view. Thus in order to contaminate the
HXD/PIN field of view, a background AGN would itself have to be heavily absorbed 
($N_{\rm H}>10^{24}$\,cm$^{-2}$) below 10\,keV 
(so that $F_{2-10}\ls 10^{-13}$\,erg\,cm$^{-2}$\,s$^{-1}$, as there are no known 
bright source in the PDS 456 field below 10 keV) and yet its spectrum must rise 
by a factor of $\times 100$ above 10 keV. 
Statistically speaking, $<5$\% of the BAT selected AGN have 
$N_{\rm H}>10^{24}$\,cm$^{-2}$, meaning that the coincidence 
probability is $\sim10^{-4}$. The only other possibility is of a flaring
object, such as a blazar, brightening during the observation, although the HXD 
lightcurve is flat (Figure 1), which appears to exclude this and also 
the object would have to be located outside the imaging field of view of the XIS. 

Furthermore, 
PDS 456 was also detected previously in both 1998 \citep{Reeves00} and 2001 with RXTE, 
with a hard X-ray flux of 
$F_{3-15}=0.7-0.9\times10^{-11}$\,erg\,cm$^{-2}$\,s$^{-1}$. 
PDS\,456 was also more recently detected in a 35\,ks worth of RXTE 
pointings in 2008 (see Table 1 for a list of observations).
PDS\,456 was not detected 
in a previous observation with \sax\ PDS (Feb 2001, 71\,ks PDS exposure), however 
the upper limit obtained is consistent with the above HXD analysis 
($F_{20-100}<1.5\times10^{-11}$\,erg\,cm$^{-2}$\,s$^{-1}$). Indeed 
although the RXTE observations and the Suzaku HXD observation are not simultaneous 
(see Table 1), the time-averaged RXTE/PCA spectrum extrapolates very well 
to the HXD/PIN measurement (see Figure 11). 
Therefore it appears likely that the detected 
hard X-ray emission is due to the quasar and not due to 
any contaminating sources in the field of view.

\clearpage

\begin{figure}
\begin{center}
\rotatebox{-90}{\includegraphics[height=12cm]{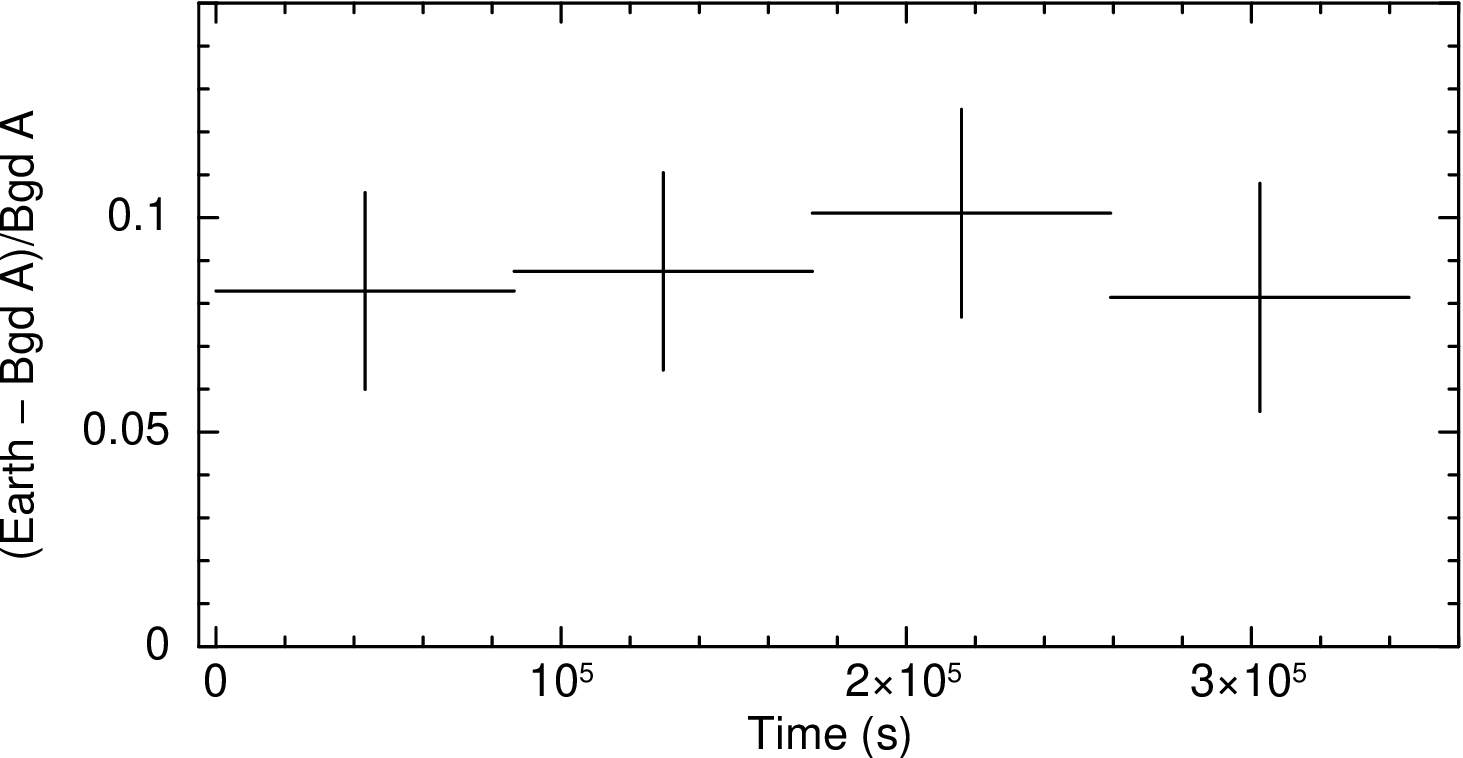}}
\rotatebox{-90}{\includegraphics[height=12cm]{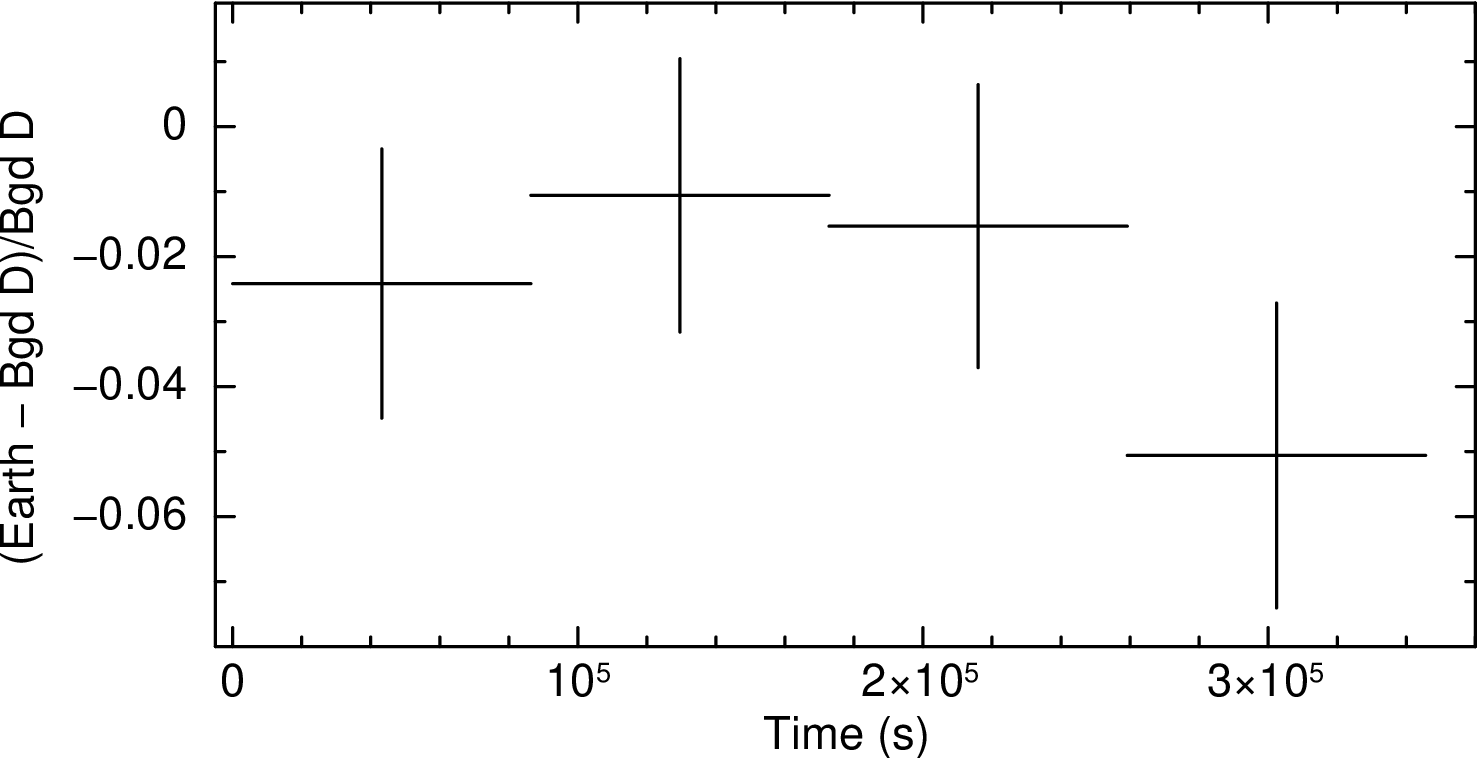}}
\rotatebox{-90}{\includegraphics[height=12cm]{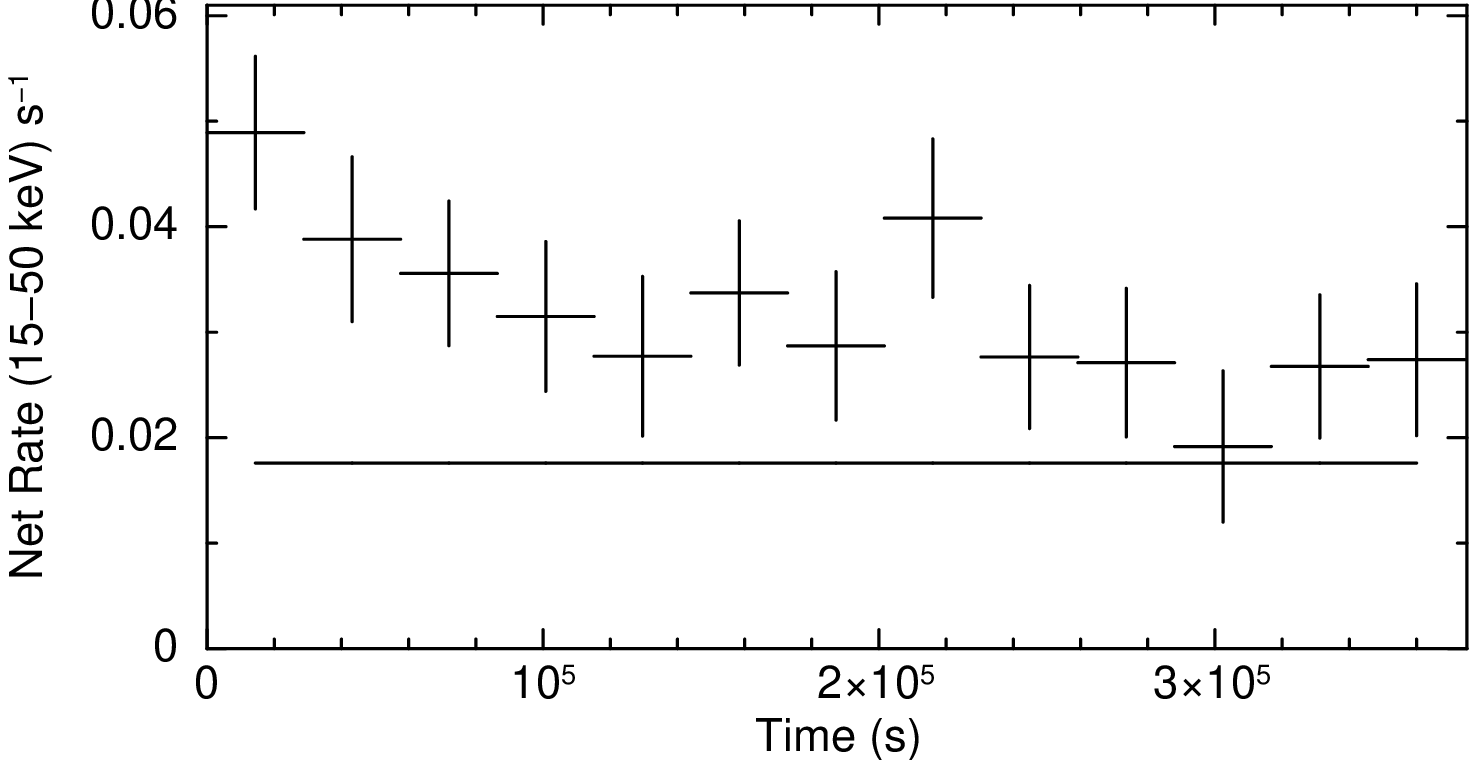}}
\end{center}
\caption{(a) Comparison between the background A model and the Earth occultation data
for HXD/PIN (from $15-50$\,keV, 
where the light curve represents the ratio between the (Earth - Bgd A) / Bgd A rate. 
The background A model underpredicts the Earth background by $8.8\pm1.0$\%. 
(b) Comparison between the background D model and the Earth occultation data, 
where the light curve represents the ratio between the (Earth $-$ Bgd D) / Bgd D rate. 
The background D model slightly underpredicts the Earth background by $2.4\pm1.0$\%. 
(c) The net 15-50 keV HXD/PIN lightcurve using the background D model, the CXB background 
level (0.17 cts\,s$^{-1}$) is marked by a horizontal line.} 
\end{figure}

\begin{figure}
\begin{center}
\rotatebox{-90}{\includegraphics[height=14cm]{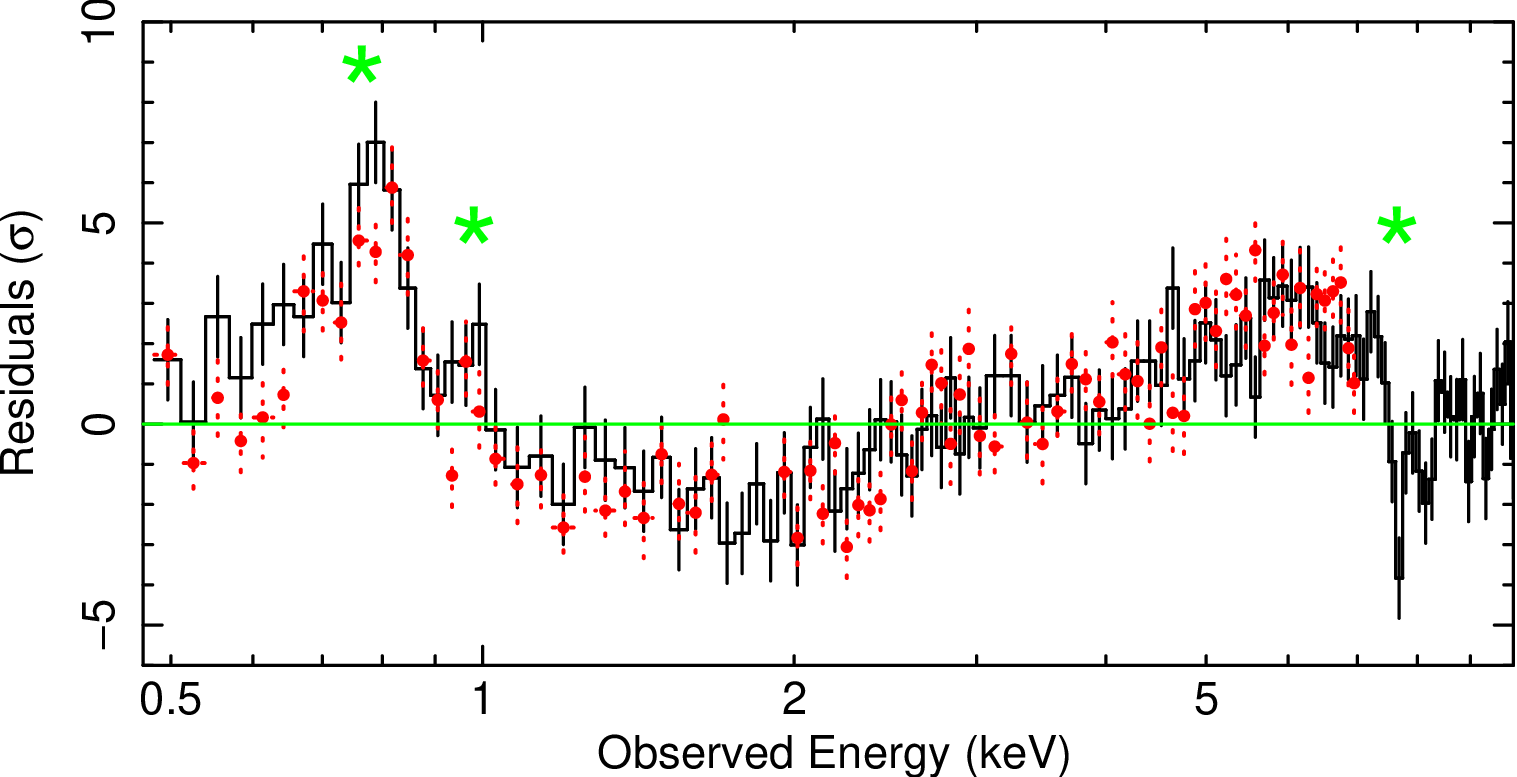}}
\rotatebox{-90}{\includegraphics[height=14cm]{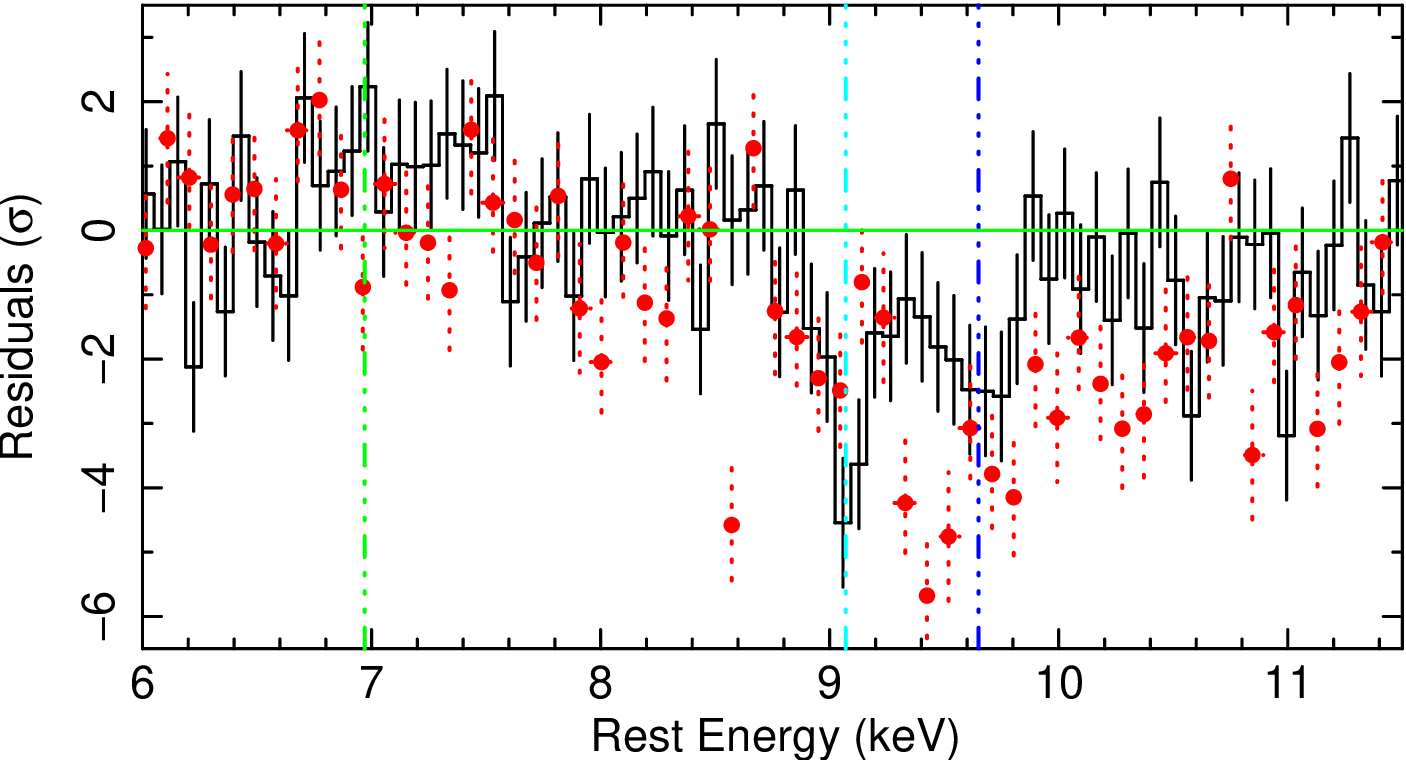}}
\end{center}
\caption{Upper panel - Suzaku XIS spectra of PDS 456, plotted as a ratio to 
a simple absorbed power-law ($\Gamma=2.2$). Note XIS\,1 is shown in red, 
XIS--FI in black. The position of the two soft X-ray emission lines and of 
the Fe K-shell absorption are marked with a star.
Lower panel - rest frame iron K-shell absorption in PDS 456 (against the 
best fit, broken power-law continuum). Suzaku XIS--FI is shown in black, 
XMM-Newton EPIC-pn (2001) in red. The green line represents the expected 
position of the Fe\,\textsc{xxvi} $1s-2p$ line (6.97 keV), while the blue lines 
represent the centroids of the detected absorption lines in XIS.} 
\end{figure}

\begin{figure}
\begin{center}
\rotatebox{-90}{
\epsscale{0.7}
\plotone{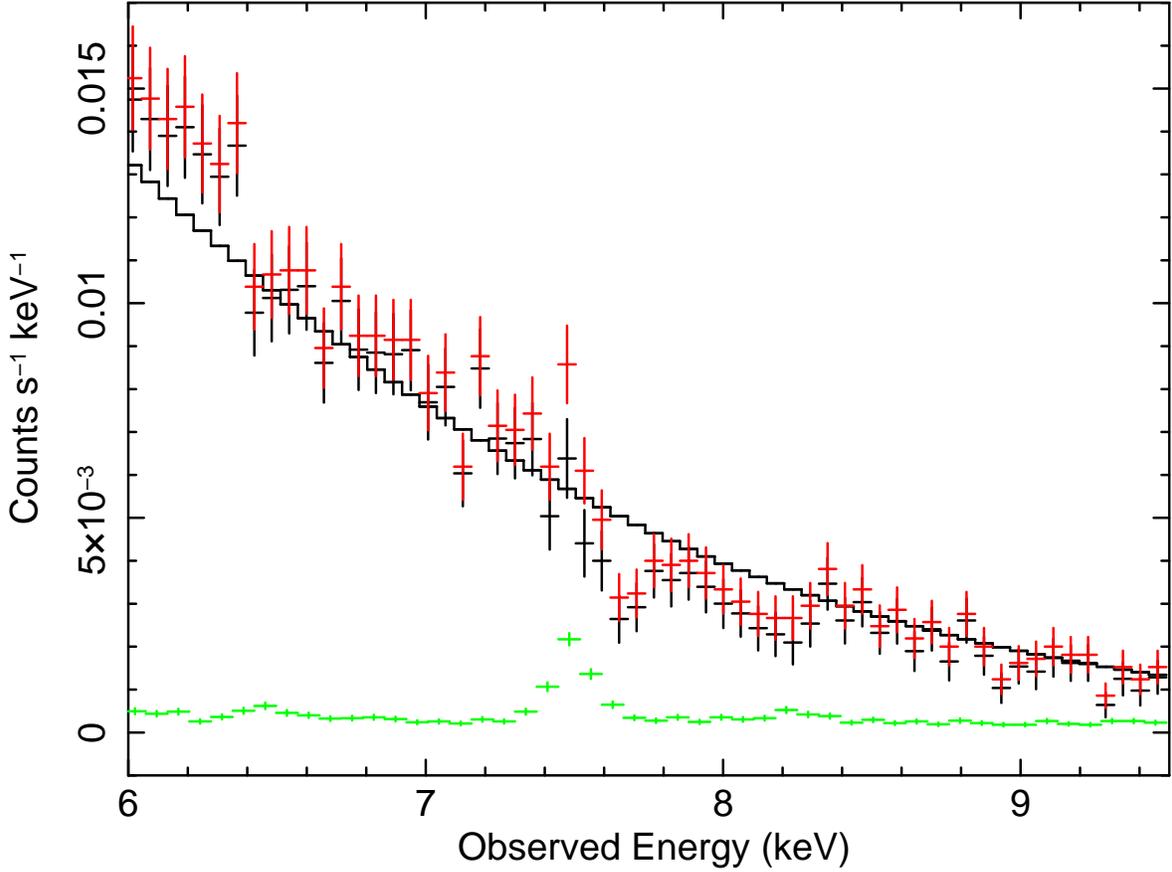}}
\caption{Suzaku XIS--FI PDS\,456 spectrum versus the background spectrum in the iron K 
band. The source spectrum prior to background subtraction in shown in red, the net 
source spectrum after background subtraction is in black and the background spectrum 
is in green. The black line shows the broken powerlaw continuum model, folded through 
the XIS response. 
It can be seen that the background subtraction has little effect in the iron K 
band and on the absorption features at 7.7 keV and 8.2 keV (observed frame). Overall the 
level of the XIS background is low and the only strong background line arises from Ni K$\alpha$ 
at 7.45\,keV, which is not coincident with the centroids of the absorption lines 
from PDS\,456. 
Note the normalization of the background has been rescaled to that of the XIS source 
region size.}
\end{center}
\end{figure}

\begin{figure}
\begin{center}
\rotatebox{-90}{\includegraphics[height=14cm]{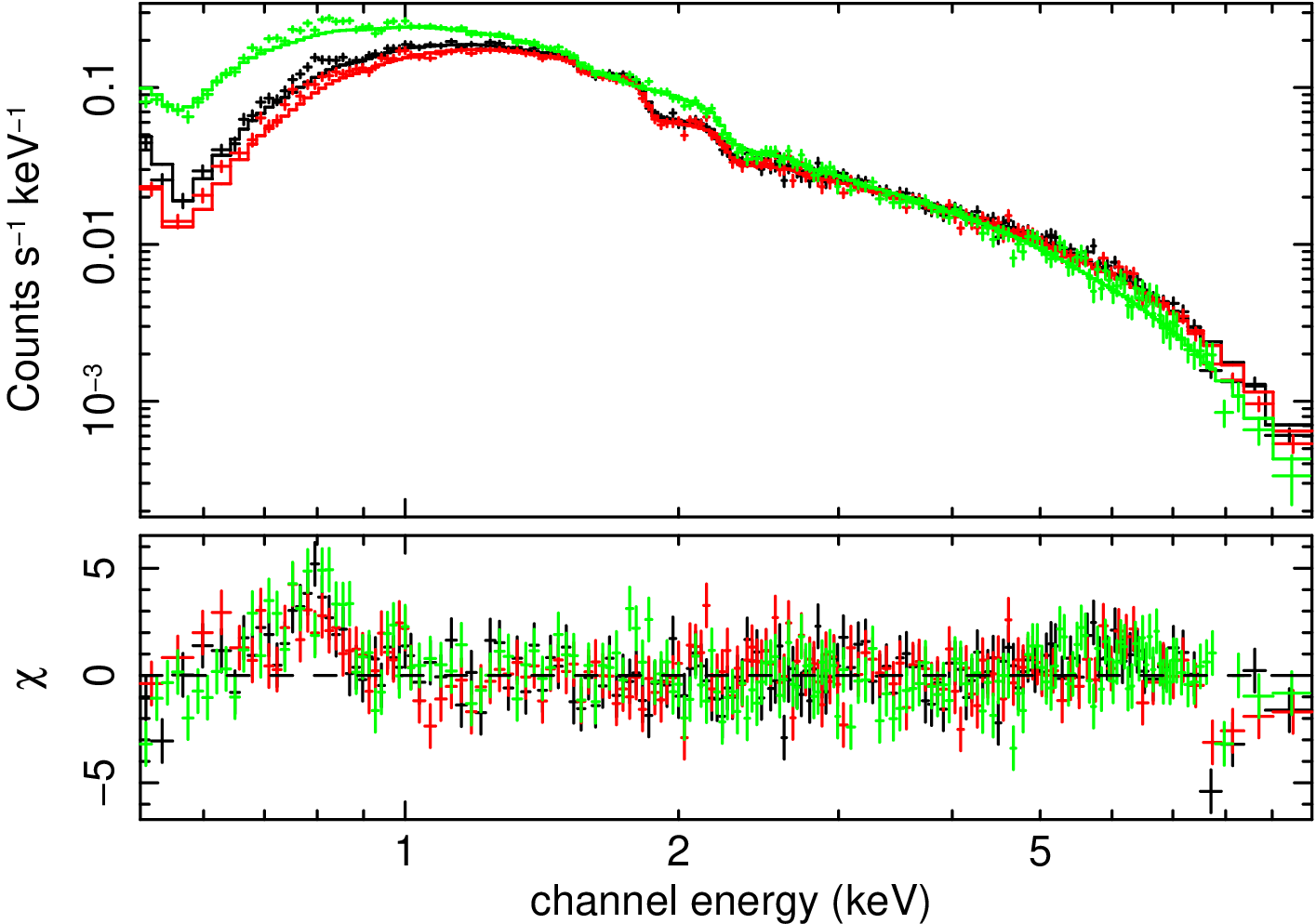}}
\rotatebox{-90}{\includegraphics[height=14cm]{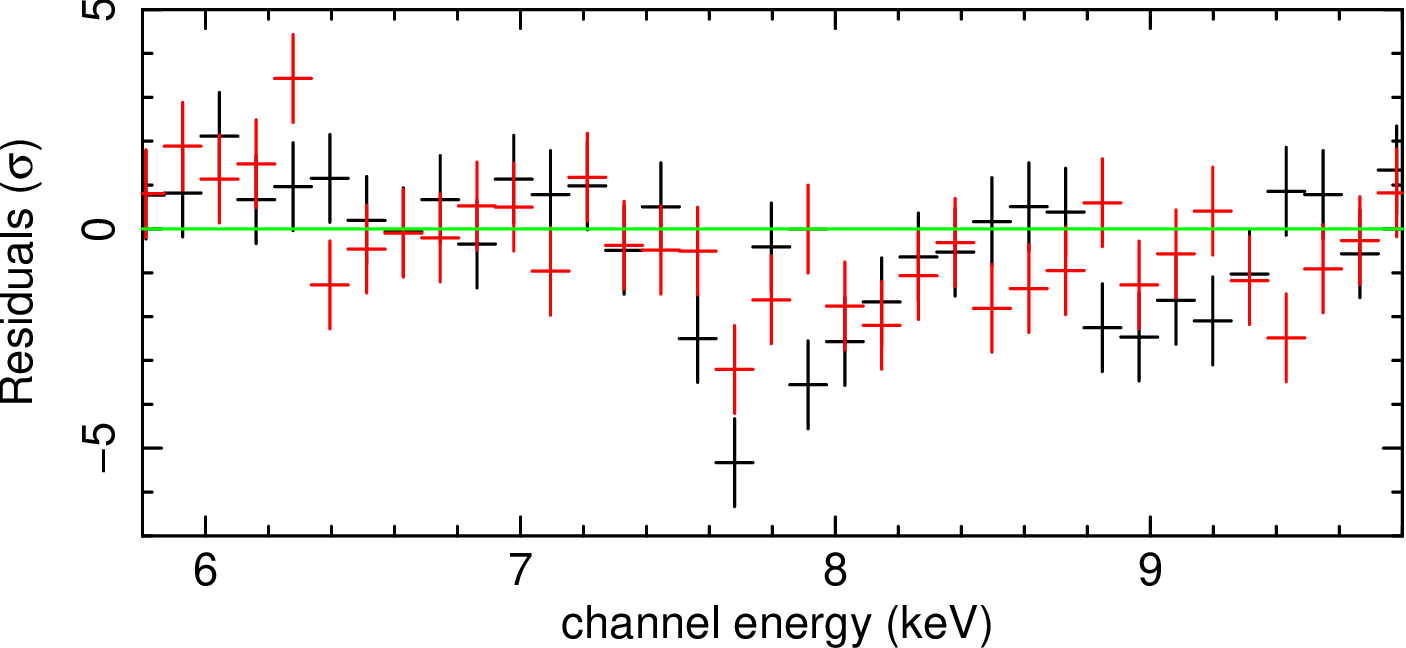}}
\end{center}
\caption{XIS 0 (black), XIS 1 (green) and XIS 3 (red) 
spectra of PDS 456, compared to the baseline broken power-law 
continuum model. The top panel shows the data folded through the instrumental response, 
the lower panel the residuals compared to the continuum. Here the data is binned 
with a minimum S/N of 10 for XIS 0, 3 and S/N of 5 for XIS 1.
Note the excess soft X-ray emission 
(0.8 keV observed) and hard X-ray absorption (7.5--8.0 keV) is present in all 3 XIS 
detectors. The lower panel show the comparison between XIS 0 and 3 in the Fe K band, 
where the iron K absorption is observed, with the data binned at the instrumental (FWHM) 
resolution.}
\end{figure}

\begin{figure}
\begin{center}
\rotatebox{-90}{\includegraphics[height=11cm]{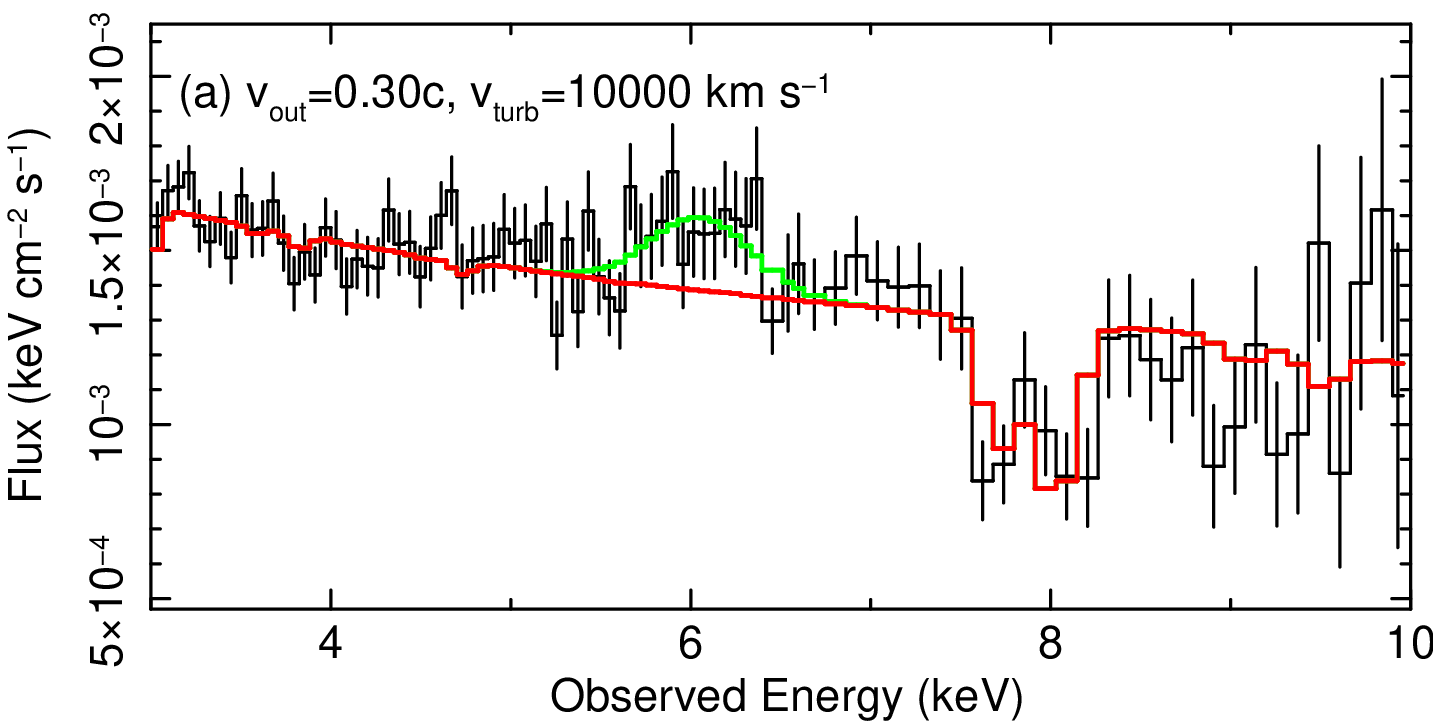}}
\rotatebox{-90}{\includegraphics[height=11cm]{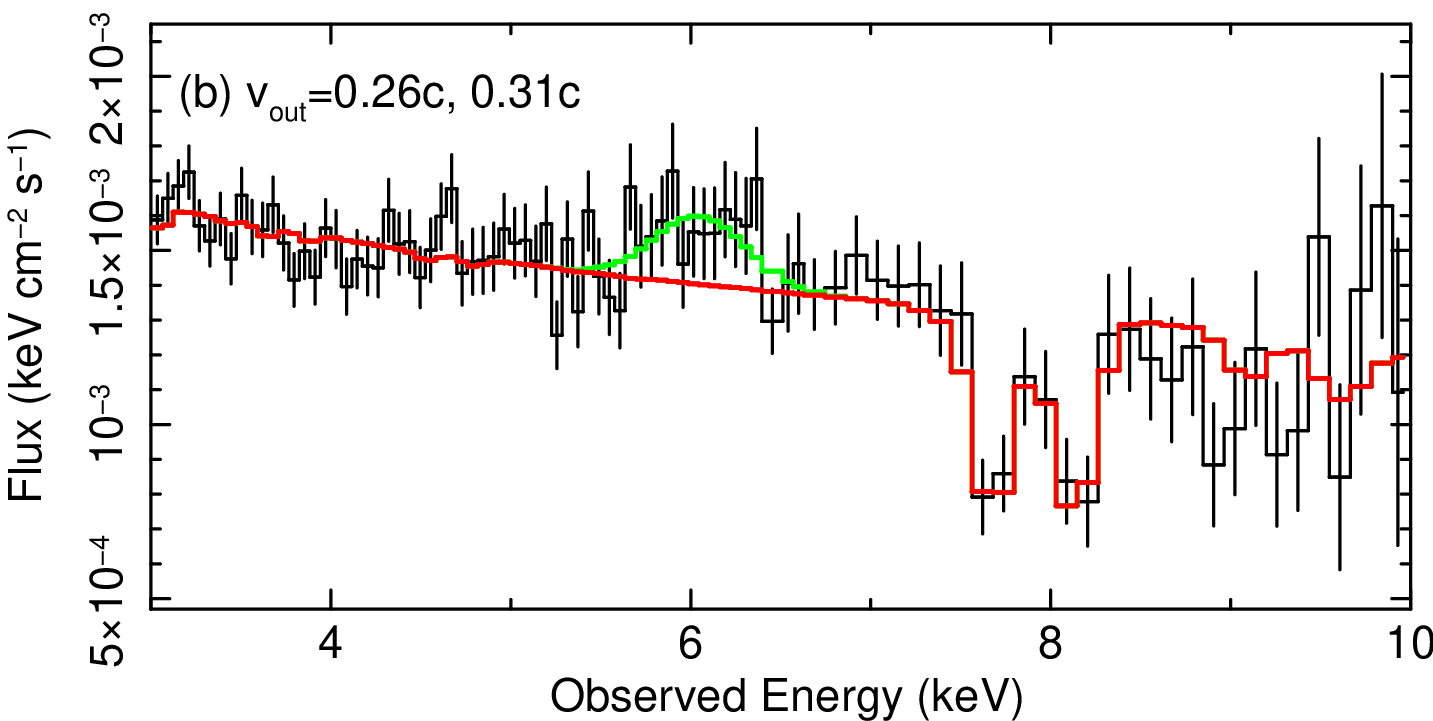}}
\rotatebox{-90}{\includegraphics[height=11cm]{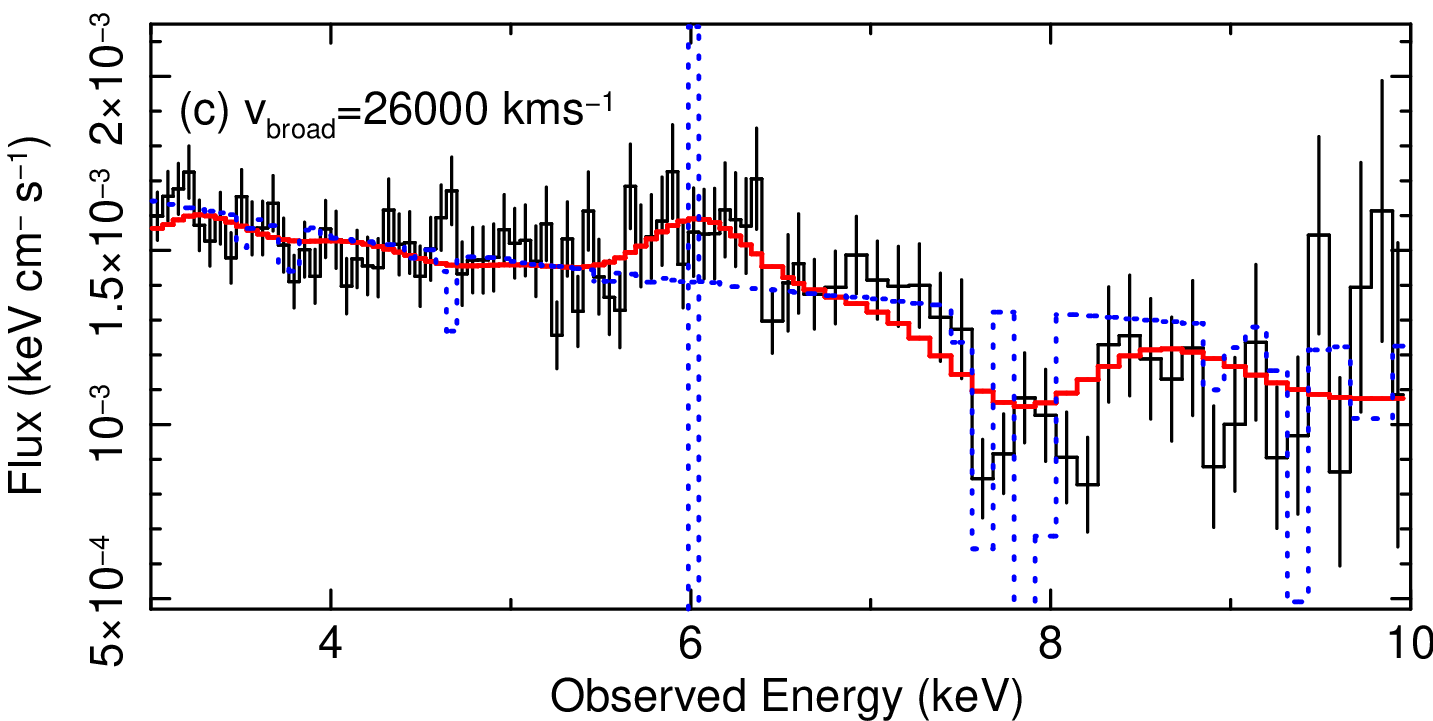}}
\end{center}
\caption{Comparison between outflowing Xstar models fitted to the Suzaku XIS--FI, showing the 
iron K band. Panel (a) shows a fit with one Xstar zone, outflowing with a single
velocity of $v_{\rm out}=-0.30c$ and assuming a turbulence velocity of
$v_{\rm turb}=10000$\,km\,s$^{-1}$ (model\,A in Section 4.2).
The absorption model is shown by the red line, the 8\,keV absorption 
(observed frame) is modeled by a blend of 
Fe\,\textsc{xxv} and Fe\,\textsc{xxv} $1s-2p$ resonance lines, 
while iron K emission is shown in green. Panel (b) shows a similar model as per (a), 
but with two outflowing zones of $v_{\rm out}=-0.26c$ and $-0.31c$ (model\,B). 
Panel (c) shows 
a similar model as per panel (a), but with Gaussian velocity broadening included 
self-consistently for the iron K and L-shell emission and absorption (model\,C).
The blue line shows the 
model without broadening. Further details are given in 
Section 4.2 and parameters for models A, B and C are listed in Table 3.}
\end{figure}

\begin{figure}
\begin{center}
\rotatebox{-90}{
\epsscale{0.5}
\plotone{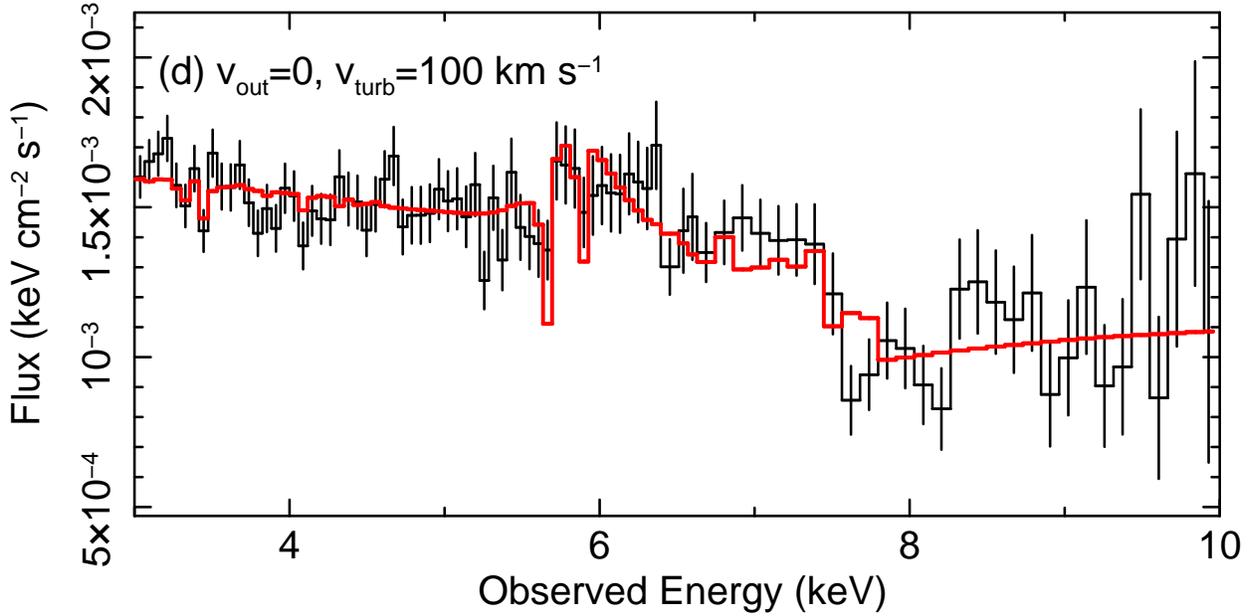}}
\caption{Xstar fit to the Suzaku XIS--FI data as per Figure 5, but with no 
net outflow velocity and assuming a very low turbulence velocity (100 km\,s$^{-1}$); 
e.g. model D in Section 4.2 and Table 3. A broad Fe K emission line is also 
included, as per models A-C in Figure 5.
Here the absorption at 8 keV in the observed frame is modeled by bound-free edges from 
Fe\,\textsc{xxv} and Fe\,\textsc{xxvi}, from a high column density absorber (i.e. $\tau=1$), 
while narrow $1s-2p$ resonance lines from Fe\,\textsc{xxv} and Fe\,\textsc{xxvi} appear 
near 6\,keV in the model (but not in the data). The fit is statistically worse 
than for the outflowing models shown in Figure 5 and such a small turbulence 
may be physically implausible.}
\end{center}
\end{figure}

\begin{figure}
\begin{center}
\rotatebox{-90}{
\epsscale{0.7}
\plotone{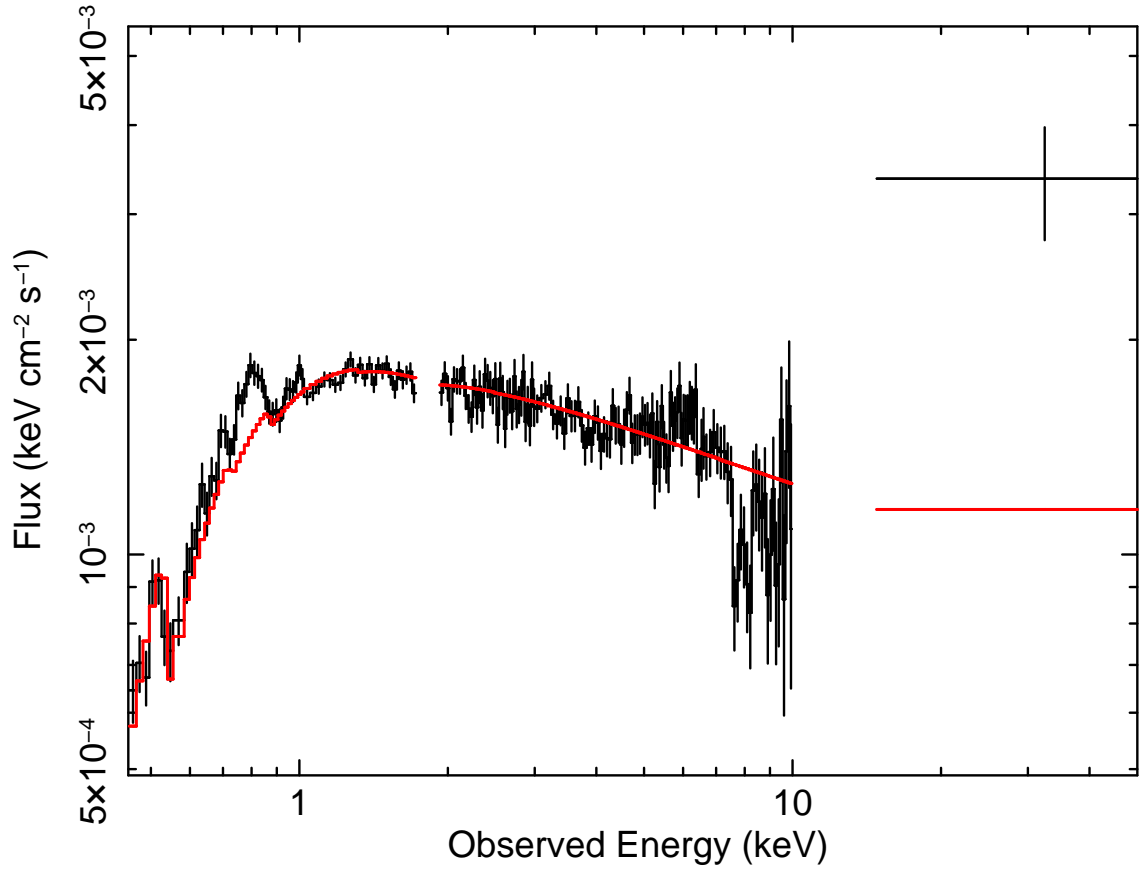}}
\caption{The Suzaku XIS and HXD spectra of PDS 456, plotted in 
$\nu F_{\nu}$ flux units. A factor of
$\times3$ excess is seen in the HXD, which may suggest that a Compton thick 
absorber, with $N_{\rm H}>10^{24}$\,cm$^{-2}$ exists towards this optical 
type I quasar. The red line shows the extrapolation of the best-fit broken-powerlaw
continuum extrapolated to higher energies.}
\end{center}
\end{figure}

\begin{figure}
\begin{center}
\rotatebox{-90}{
\epsscale{0.7}
\plotone{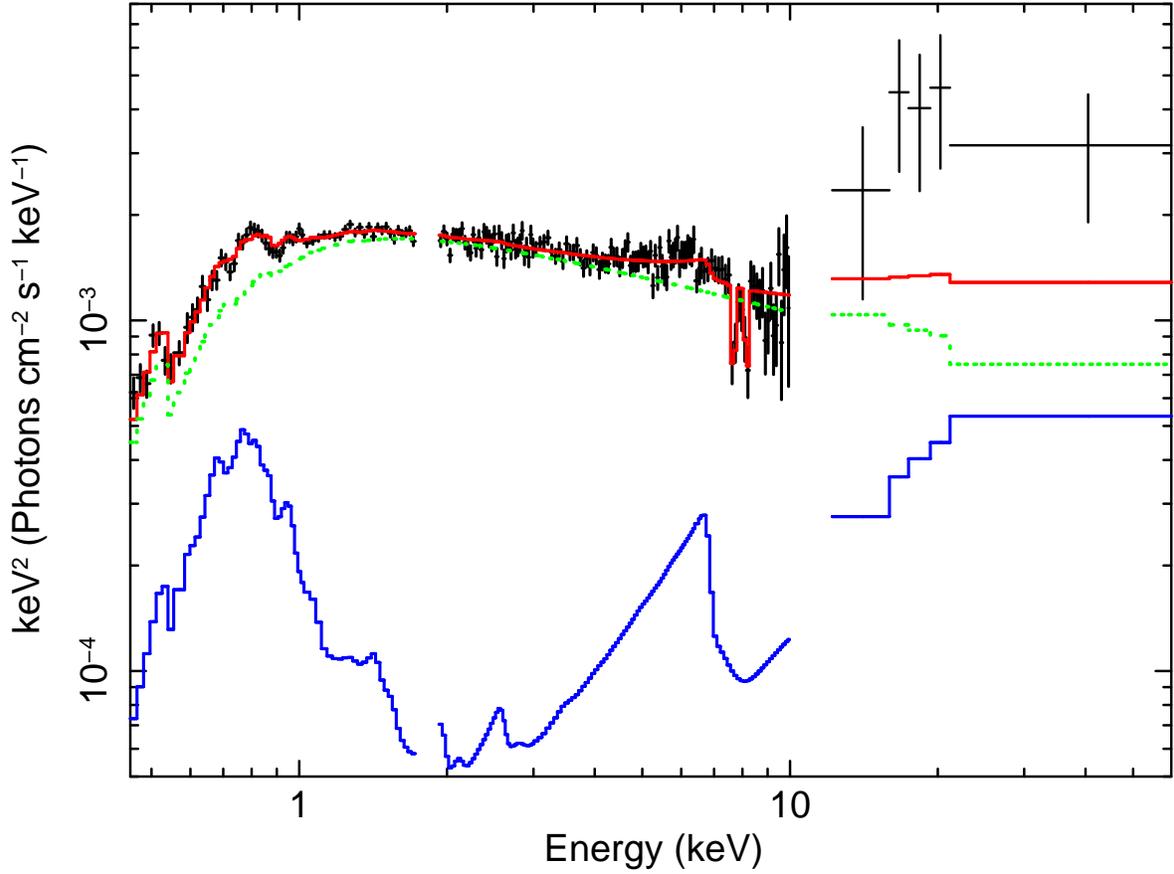}}
\caption{Suzaku XIS+HXD spectrum of PDS 456, fitted with an ionized disk 
reflection model, as described in Section 5.1. 
Data are shown in black, the blue line shows the blurred 
reflection component, green shows the intrinsic power-law continuum and red the 
total emission.  Note the HXD/PIN spectrum is more finely binned than in 
Figure 7, at $\sim 2\sigma$ per bin above the background level.}
\end{center}
\end{figure}

\begin{figure}
\begin{center}
\rotatebox{270}{\resizebox{12cm}{16cm}{\includegraphics{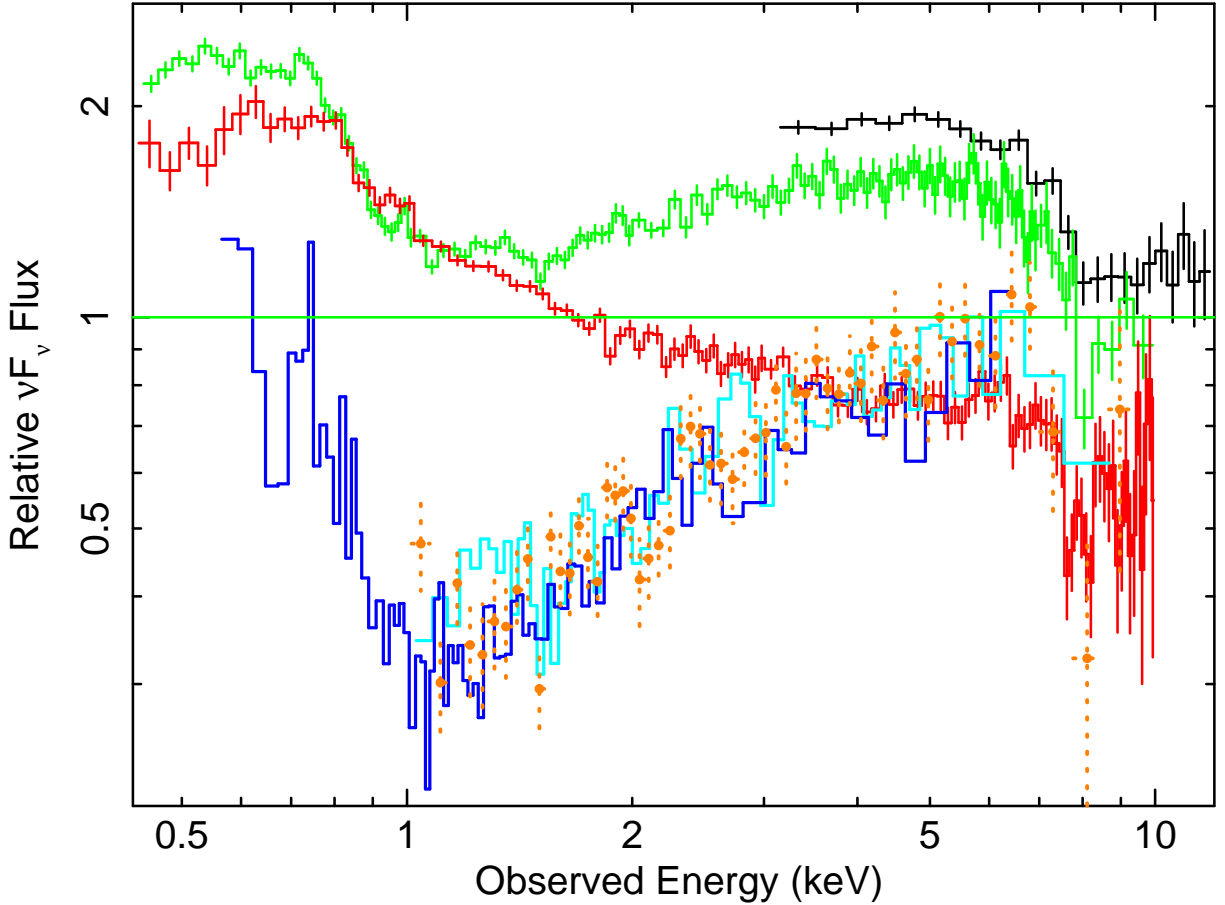}}}
\caption{The current observations of PDS 456, shown as $\nu F_{\nu}$ 
spectra, where we have plotted the data compared to a ratio with an absorbed 
($N_{\rm Gal}=2\times10^{21}$\,cm$^{-2}$) power-law of $\Gamma=2$ at 
a mean 2--10\,keV flux level of $5\times 10^{-12}$\,erg\,cm$^{-2}$\,s$^{-1}$.  
Shown are RXTE PCA (black, 1998), XMM-Newton (2001, green), Suzaku XIS (2007, red), 
Chandra/HETG (2003, blue) and ASCA (1998, orange dotted).  
Strong variability is seen below 10 keV, from RXTE and XMM  at high fluxes, 
down to Chandra and ASCA as the lowest/hardest observations. Note the 8 keV (observed) 
Fe K-shell absorption in the XMM--2001, Suzaku--2007 and RXTE--1998 observations. 
A summary of all these observations is shown in Table 1.}
\end{center}
\end{figure}

\begin{figure}
\begin{center}
\rotatebox{-90}{
\epsscale{0.7}
\plotone{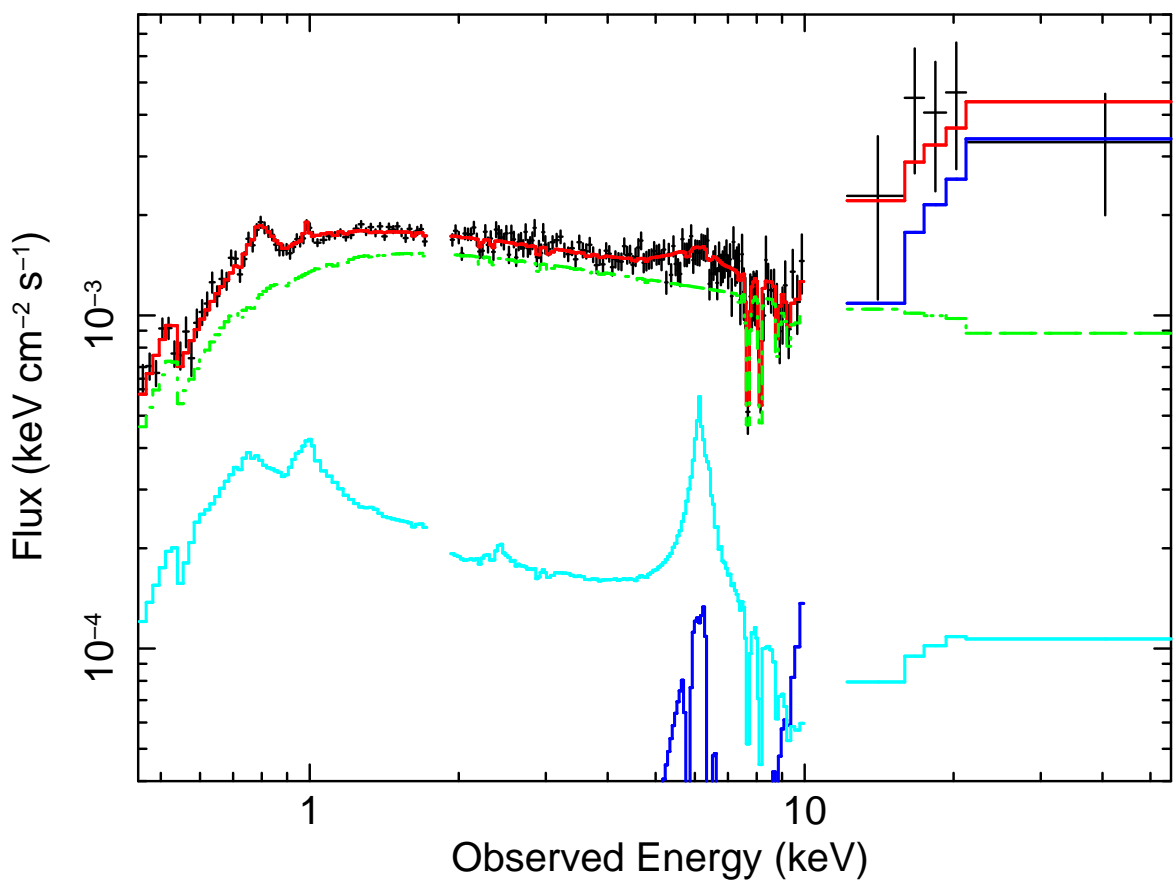}}
\caption{Model fitted to the Suzaku spectrum, representative of absorption and emission 
from a high column density disk wind, as described in Section 6.3, with model 
parameters listed in Table 4. Datapoints are shown in black, 
the dotted green line represents the power-law ($\Gamma=2.2)$ continuum that is absorbed 
only by the very high ionization fast outflowing zone, the dark blue line is the continuum 
obscured by the partial covering absorber which covers 80\% of the X-ray source, 
the light-blue line is the reflected emission off the Compton-thick 
partially ionized wind and red is the total emission from all components.}
\end{center}
\end{figure}

\begin{figure}
\begin{center}
\rotatebox{-90}{
\epsscale{0.7}
\plotone{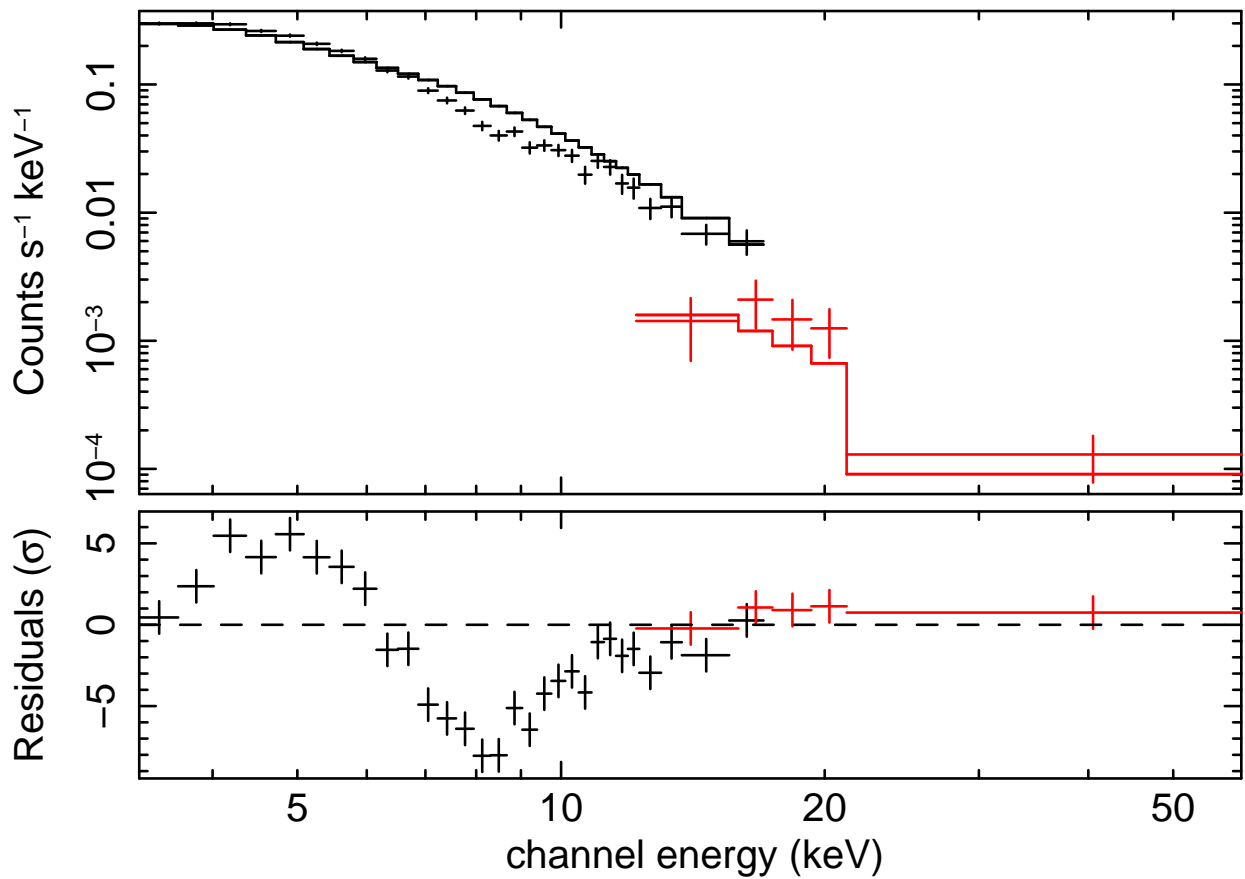}}
\caption{A comparison between the time--averaged RXTE/PCA spectrum between 3--18\,keV 
and the HXD/PIN Suzaku spectrum from 15--50\,keV. The plot shows the comparison 
of the data to a $\Gamma=2.2$ continuum with Galactic absorption, with no re-normalization 
of either spectrum, while the lower panel shows the residuals between the data and the model. 
Although the RXTE/PCA 
data is non simultaneous with the HXD (see Table 1 for a list of observations), 
the PCA spectrum below 
18\,keV extrapolates well to the HXD/PIN data above 15\,keV. Also notice the deep  
absorption trough centered at 8 keV in the RXTE/PCA spectrum, consistent with the 
absorption in the 2007 Suzaku XIS and 2001 XMM-Newton data.}
\end{center}
\end{figure}

\clearpage

\begin{deluxetable}{lccccc}
\tabletypesize{\small}
\tablecaption{Summary of PDS 456 Observations}
\tablewidth{0pt}
\tablehead{
\colhead{Telescope} & \colhead{Mode$^{a}$} & \colhead{Start Date/Time$^{b}$} 
& \colhead{End Date/Time$^{b}$} & \colhead{Exposure$^{c}$} 
& \colhead{$F_{2-10}$$^{d}$}}

\startdata

Suzaku XIS & XIS nom & 2007/02/24 17:58:04 & 2007/03/01 00:51:14 & 179.6 & 3.7 \\
Suzaku HXD/PIN & -- & -- & -- & 164.8 & -- \\
XMM-Newton pn & FW/med & 2001/02/26 10:46:57 & 2001/02/26 22:00:41 & 36.4 & 6.9 \\
Chandra/HETG & & 2003/05/07 03:29:56 & 2003/05/08 20:08:33 & 142.8 & 3.8 \\
ASCA/SIS & & 1998/03/07 09:06:36 & 1998/03/08 14:58:42 & 42.8 & 4.0 \\
RXTE/PCA & & 1998/03/07 17:07:44 & 1998/03/10 07:50:56 & 92.1 & 8.1 \\
         & & 2001/02/23 04:20:32 & 2001/03/11 05:59:44 & 168.3 & 10.4 \\
         & & 2008/02/22 21:00:32 & 2008/11/14 08:17:36 & 34.7 & 6.9 \\
Beppo-SAX/PDS & & 2001/02/26 09:50:03 & 2001/03/03 08:57:30 & 71.0 & -- \\

\enddata


\tablenotetext{a}{Instrument mode or pointing. XIS nom = XIS nominal pointing;  
FW = Full Window; med = medium filter.} 
\tablenotetext{b}{Observation Start/End times are in UT.} 
\tablenotetext{c}{Net exposure time, after screening and deadtime correction, in ks.}
\tablenotetext{d}{Absorbed flux in the 2-10\,keV band, units $\times10^{-12}$\,erg\,cm$^{-2}$\,s$^{-1}$}

\end{deluxetable}

\clearpage

\begin{deluxetable}{lccc}
\tabletypesize{\small}
\tablecaption{Spectral Parameters to Suzaku XIS}
\tablewidth{0pt}
\tablehead{
\colhead{Model Component} & \colhead{Fit Parameter} & \colhead{Value} &
\colhead{$\Delta \chi^{2}$$^{a}$} }

\startdata

1. Broken Power-law Continuum$^{b}$ & $\Gamma_{\rm soft}$ & $2.44\pm0.04$ \\

& $E_{\rm break}$ & $2.0\pm0.2$ \\

& $\Gamma_{\rm hard}$ & $2.24\pm0.02$ \\

& $N_{\rm BPL}$ & $(2.4\pm0.1)\times10^{-3}$ \\

& $F_{0.5-10 keV}$ & $7.2\times10^{-12}$ \\


2. Galactic Absorption$^{c}$ & $N_{\rm H}$ & $(2.2\pm0.2)\times10^{21}$ \\

3. Soft X-ray Emission Lines$^{d}$ & $E_{\rm line}$ & $0.91\pm0.01$ & 174.4 \\

& Line Flux & $(2.0\pm0.4)\times10^{-4}$ \\

& EW & $42\pm10$ \\

& $\sigma$ & $0.075\pm0.015$ \\

Second line & $E_{\rm line}$ & $1.16\pm0.02$ & 17.7 \\

& Line Flux & $(1.4\pm0.6)\times10^{-5}$ \\

& EW & $6\pm3$ \\

& $\sigma$ & $0.01^{f}$ \\

4. Iron K Absorption Lines$^{d}$ & $E_{\rm line}$ & $9.09\pm0.05$ & 41.1 \\

& Line Flux & $-3.0^{+1.5}_{-1.1}\times10^{-6}$ \\

& EW & $133\pm39$ \\

Second line & $E_{\rm line}$ & $9.64\pm0.08$ & 29.0\\

& Line Flux & $-(2.5\pm1.0)\times10^{-6}$ \\

& EW & $129\pm48$ \\

& $\sigma$ & $0.105^{+0.055}_{-0.065}$ (tied to first line) \\

5. Iron K Emission Line$^{d}$ & $E_{\rm line}$ & $7.1\pm0.2$ & 27.7 \\

& Line Flux & $(5.1\pm2.0)\times10^{-6}$ \\

& EW & $133\pm52$\,eV \\

& $\sigma$ & $0.35^{+0.65}_{-0.14}$ \\

Fit statistics$^{e}$ & $\chi^{2}/{\rm dof}$ & $431.5/374$ \\

& Null probability & $2.1\times10^{-2}$ \\

\enddata


\tablenotetext{a}{Improvement in $\chi^{2}$ to fit after adding model component.}
\tablenotetext{b}{Broken power-law continuum parameters. 
$\Gamma_{\rm hard}$ and $\Gamma_{\rm soft}$, soft and hard photon indicies; $E_{\rm break}$, 
break energy in units keV; $N_{\rm BPL}$, broken powerlaw normalization in units 
photons\,cm$^{-2}$\,s$^{-1}$ at 1\,keV; $F_{0.5-10 keV}$, absorbed continuum flux 
in units ergs\,cm$^{-2}$\,s$^{-1}$.}
\tablenotetext{c}{Local Galactic absorption (at $z=0$), units cm$^{-2}$.}
\tablenotetext{d}{Emission/Absorption line parameters. $E_{\rm line}$, line energy in units 
keV; line flux in units photons\,cm$^{-2}$\,s$^{-1}$; equivalent width (EW) in units eV; 
$\sigma$ is $1\sigma$ line width in units keV. Parameters given in quasar rest-frame.}
\tablenotetext{e}{Reduced chi-squared ($\chi^{2}/{\rm dof}$)
and null hypothesis probability for spectral fit.}
\tablenotetext{f}{Parameter is fixed in spectral fit.}

\end{deluxetable}

\clearpage

\begin{deluxetable}{lccccccc}
\tabletypesize{\small}
\tablecaption{Xstar Photoionization Model Fits to Suzaku XIS}
\tablewidth{0pt}
\tablehead{
\colhead{Model$^{a}$} & \colhead{$N_{H}$$^{b}$} & \colhead{log\,$\xi$$^{c}$} 
& \colhead{$\sigma_{\rm turb}$$^{d}$} & \colhead{$v_{\rm out}$$^{e}$} & 
\colhead{$v_{\rm broad}$$^{f}$} & \colhead{$\chi^{2}/{\rm dof}$$^{g}$} & 
\colhead{Null Prob$^{g}$}\\
}
\startdata

A & $0.22^{+0.11}_{-0.10}$ & $4.1\pm0.2$ & 10000 & $-0.30\pm0.02c$ & -- & 224.9/204 & 0.151 \\ 

B & $1.3\pm0.6$ & $4.9^{+0.9}_{-0.5}$ & 10000 & $-0.26\pm0.02c$ & -- & 207.3/203 & 0.394 \\

& tied & tied & tied & $-0.31\pm0.02c$ \\

C & $1.5\pm0.6$ & $4.4\pm0.4$ & 10000 & $-0.29\pm0.02c$ & $26000\pm7000$ 
& 227.9/205 & 0.131 \\

D & $1.2\pm0.3$ & $4.0\pm0.2$ & 100 & 0 & -- & 233.3/205 & $8.5\times10^{-2}$ \\

E & $1.3\pm0.5$ & $4.0\pm0.2$ & 100 & 0 & -- & 228.3/204 & 0.118 \\

\enddata


\tablenotetext{a}{Models A, B, C, D and E as defined in the text (section 4.2). 
Model A contains a single photoionized absorber with one outflow velocity; model B 
contains two absorbers with two seperate outflow velocities; model C contains one outflow 
velocity with equal kinematical broadening in emission/absorption; 
model D has zero outflow velocity and low ($100$\,km\,s$^{-1}$) turbulence; model E as 
per model D, but allowing for 2 additional narrow iron emission lines at 6.70\,keV and 
6.97\,keV rest frame.}
\tablenotetext{b}{Column density, in units $\times10^{24}$\,cm$^{-2}$.}
\tablenotetext{c}{Ionization parameter, units ergs\,cm\,s$^{-1}$.}
\tablenotetext{d}{Turbulence velocity width ($1\sigma$), in units km\,s$^{-1}$.}
\tablenotetext{e}{Outflow velocity, in units of $c$.}
\tablenotetext{f}{Gaussian velocity broadening (FWHM), in units km\,s$^{-1}$.}
\tablenotetext{g}{Reduced chi-squared ($\chi^{2}/{\rm dof}$)
and null hypothesis probability for spectral fit.}

\end{deluxetable}

\clearpage

\begin{deluxetable}{lcc}
\tabletypesize{\small}
\tablecaption{Outflowing Wind Model Parameters}
\tablewidth{0pt}
\tablehead{
\colhead{Model Component} & \colhead{Fit Parameter} & \colhead{Value} \\
}
\startdata

1. Power-law Continuum & $\Gamma$ & $2.25\pm0.05$ \\

& $N_{\rm abs}$$^{a}$ & $(8.1\pm2.3)\times10^{-3}$ \\

& $N_{\rm unabs}$$^{a}$ & $(2.0\pm0.1)\times10^{-3}$ \\

& $F_{0.5-10 keV}$$^{b}$ & $7.2\times10^{-12}$ \\

& $F_{15-50 keV}$$^{b}$ & $7.8\times10^{-12}$ \\

2. Outflowing Fe K-shell Absorber & $N_{\rm H}$$^{c}$ & $(1.3\pm0.5)\times10^{24}$ \\

& log\,$\xi$$^{d}$ & $4.9^{+0.9}_{-0.5}$ \\

& $v_{\rm out1}$$^{e}$ & $-0.26\pm0.02c$ \\

& $v_{\rm out2}$$^{e}$ & $-0.31\pm0.02c$ \\

3. Compton-thick Hard X-ray Absorber & $N_{\rm H}$$^{c}$ & $>2.5\times10^{24}$ \\

& log\,$\xi$$^{d}$ & $2.5\pm0.4$ \\

& $f_{\rm cov}$$^{f}$ & $0.80^{+0.04}_{-0.07}$ \\

4. Ionized Reflection & log\,$\xi$$^{d}$ & $3.0\pm0.3$ \\

& $v_{\rm broad}$$^{g}$ & $35000\pm17000$ \\

& $v_{\rm refl}$$^{h}$ & $-38000\pm8000$ \\

& $R$$^{i}$ & $0.2$ \\

Fit statistics & $\chi^{2}/{\rm dof}$$^{j}$ & $217.3/211$ \\

& Null probability$^{j}$ & $0.367$ \\

\enddata


\tablenotetext{a}{Powerlaw normalization in units photons\,cm$^{-2}$\,s$^{-1}$ 
at 1\,keV; $F_{0.5-10 keV}$. $N_{\rm abs}$ refers to the power-law component that is absorbed 
by the hard X-ray absorber, $N_{\rm unabs}$ is the component not covered 
by the hard X-ray absorber.}
\tablenotetext{b}{Measured flux in units ergs\,cm$^{-2}$\,s$^{-1}$.}
\tablenotetext{c}{Column density, units cm$^{-2}$.}
\tablenotetext{d}{Ionization parameter, units ergs\,cm\,s$^{-1}$.}
\tablenotetext{e}{Iron K absorber outflow velocity, in units of $c$.}
\tablenotetext{f}{Absorber covering fraction}
\tablenotetext{g}{FWHM Gaussian velocity broadening in units km\,s$^{-1}$.}
\tablenotetext{h}{Net outflow velocity of reflector in units km\,s$^{-1}$.}
\tablenotetext{i}{Ratio of reflected emission to power-law continuum.}
\tablenotetext{j}{Reduced chi-squared ($\chi^{2}/{\rm dof}$)
and null hypothesis probability for spectral fit.}

\end{deluxetable}

\clearpage


\clearpage

\begin{thebibliography}{}

\bibitem[Antonucci(1993)]{Ant93}
Antonucci, A. 1993, ARA\&A, 31, 473

\bibitem[Arnaud \& Rothenflug(1985)]{Arnaud85} 
Arnaud, M., \& Rothenflug, R., 1985, A\&AS, 60, 425

\bibitem[Behar et al.(2009)]{Behar09}
Behar, E., Kaspi, S., Reeves, J.N., Turner, T.J., Mushotzky, R., \& O'Brien, P.T., 2009, 
ApJ, submitted

\bibitem[Behar et al.(2003)]{Behar03} 
Behar,~E., Rasmussen,~A.~P., Blustin,~A.~J., Sako,~M., Kahn,~S.~M.,
Kaastra,~J.~S., Branduardi-Raymont,~G., \& Steenbrugge,~K.~C. 2003, \apj,
598, 232

\bibitem[Blustin et al.(2005)]{Blustin05}
Blustin, A. J., Page, M. J., Fuerst, S. V., Branduardi--Raymont, G., \& 
Ashton, C. E., 2005, A\&A, 431, 111

\bibitem[Braito et al.(2007)]{Braito07} 
Braito, V., et al., 2007, ApJ, 670, 978

\bibitem[Cappi(2006)]{Cappi06}
Cappi, M., 2006, AN, 327, 101

\bibitem[Chartas et al.(2003)]{Chartas03} 
Chartas, G., Brandt, W. N., \& Gallagher, S. C., 2003, ApJ, 595, 85

\bibitem[Chartas et al.(2002)]{Chartas02} 
Chartas, G., Brandt, W. N., Gallagher, S. C., \& Garmire, G. P., 2002, ApJ, 579, 169 

\bibitem[Churazov et al.(2007)]{Chur07}
Churazov, E. et al., 2007, A\&A, 467, 529 

\bibitem[Dadina et al.(2005)]{Dadina05}
Dadina, M., Cappi, M., Malaguti, G., Ponti, G., \& De Rosa, A. 
2005, A\&A, 442, 461

\bibitem[Dickey \& Lockman(1990)]{DL90}
Dickey, J. M. \& Lockman, F. J., 1990, ARA\&A, 28, 215

\bibitem[Di Matteo et al.(2005)]{DiMatteo05}
Di Matteo, T., Springel, V., \& Hernquist, L., 2005, Nature, 433, 604


\bibitem[Elvis et al.(1994)]{Elvis94}
Elvis, M., et al., 1994, ApJS, 95, 1

\bibitem[Fabian(1999)]{Fabian99}
Fabian, A. C., 1999, MNRAS, 308, L39

\bibitem[Fabian et al.(1989)]{Fabian89}
Fabian, A. C., Rees, M. J., Stella, L., \& White, N.E. 1989, MNRAS, 238, 729

\bibitem[Ferrarese \& Merritt(2000)]{FM00}
Ferrarese, L., \& Merritt, D., 2000, ApJ, 539, L9

\bibitem[Frontera et al.(2007)]{Frontera07}
Frontera, F., et al., 2007, ApJ, 666, 86

\bibitem[Fukazawa et al.(2009)]{Fukazawa09}
Fukazawa, Y., et al., 2009, PASJ, 61, S17

\bibitem[Gebhardt et al.(2000)]{Gebhardt00}
Gebhardt, K., et al., 2000, ApJ, 539, L13

\bibitem[George \& Fabian(1991)]{GF91}
George, I. M., \& Fabian, A. C. 1991, MNRAS, 249, 352


\bibitem[Gibson et al.(2005)]{Gibson05}
Gibson, R. R., Marshall, H. L., Canizares, C. R., \& Lee, J. C., 
2005, ApJ, 627, 83

\bibitem[Grevesse \& Sauval(1998)]{GS98} 
Grevesse, N., \& Sauval, A. J., 1998, Space Sci. Rev., 85, 161

\bibitem[Gruber et al.(1999)]{Gruber99}
Gruber, D. E., Matteson, J. L., Peterson, L. E., \& Jung, G. V. 1999,
ApJ, 520, 124 

\bibitem[Hasinger et al.(2002)]{Hasinger02} 
Hasinger,~G., Schartel,~N. \& Komossa,~S., 2002, \apj, 573, L77

\bibitem[Kalberla et al.(2005)]{Kalberla05}
Kalberla, P. M. W., Burton, W. B., Hartmann, D., Arnal, E. M., Bajaja, E., 
Morras, R., \& Pöppel, W. G. L., 2005, A\&A, 440, 775

\bibitem[Kallman et al.(2004)]{Kallman04} 
Kallman, T. R., Palmeri, P., Bautista, M. A., Mendoza, C., 
\& Krolik, J. H., 2004, ApJS, 155, 675

\bibitem[Kaspi et al.(2000)]{Kaspi00} 
Kaspi, S., Smith, P. S., Netzer, H., Maoz, D., Jannuzi, B. T., \& Giveon, U.,
2000, ApJ, 533, 631

\bibitem[Kato et al.(2004)]{Kato04}
Kato, Y., Mineshige, S., \& Shibata, K., 2004, ApJ, 605, 307

\bibitem[King \& Pounds(2003)]{KP03} 
King, A. R., \& Pounds, K. A., 2003, MNRAS, 345, 657

\bibitem[King(2003)]{King03} 
King, A. R., 2003, ApJ, 596, L27

\bibitem[Koyama et al.(2007)]{Koyama07} 
Koyama, K., et al., 2007, PASJ, 59, 23

\bibitem[Krivonos et al.(2007a)]{Kriv07a}
Krivonos, R., Revnivtsev, M., Lutovinov, A., Sazonov, S., 
Churazov, E., \& Sunyaev, R., 2007, A\&A, 475, 775

\bibitem[Krivonos et al.(2007b)]{Kriv07b}
Krivonos, R.; Revnivtsev, M., Churazov, E., Sazonov, S., 
Grebenev, S., Sunyaev, R., 2007, A\&A, 463, 957

\bibitem[Laor(1991)]{Laor91}
Laor, A. 1991, ApJ, 376, L90

\bibitem[Lightman \& White(1988)]{LW88}
Lightman, A. P., \& White, T. R. 1988, ApJ, 335, L57

\bibitem[Markowitz et al.(2006)]{Markowitz06} 
Markowitz, A., Reeves, J. N., \& Braito, V., 2006, ApJ, 646, 783

\bibitem[McKernan et al.(2005)]{McKernan05} 
McKernan, B., Yaqoob, T., \& Reynolds, C. S., 2005, MNRAS, 361, 1337

\bibitem[McKernan et al.(2004)]{McKernan04} 
McKernan, B., Yaqoob, T., \& Reynolds, C. S., 2004, ApJ, 617, 232

\bibitem[McLure \& Jarvis(2002)]{MJ02}
McLure, R. J., Jarvis, M. J., 2002, MNRAS, 337, 109

\bibitem[Mitsuda et al.(2007)]{Mitsuda07} 
Mitsuda, K., et al., 2007, PASJ, 59, 1

\bibitem[O'Brien et al.(2005)]{O'Brien05}
O'Brien, P. T., Reeves, J. N., Simpson, C., Ward, M. J., 2005, MNRAS, 360, L25

\bibitem[Papadakis et al.(2007)]{Papadakis07}
Papadakis, I. E., Brinkmann, W., Page, M. J., McHardy, I., \& Uttley, P., 
2007, A\&A, 461, 931

\bibitem[Porquet et al.(2004)]{Porquet04} 
Porquet, D., Reeves, J.N., Uttley, P., Turner, T.J.\ 2004, A\&A, 427, 101

\bibitem[Pounds \& Reeves(2009)]{PR09} 
Pounds, K. A., \& Reeves, J. N., 2009, MNRAS, accepted (arXiv:0811.3108v2)

\bibitem[Pounds et al.(2003)]{Pounds03} 
Pounds, K. A., Reeves, J. N., King, A. R., Page, K. L., 
O'Brien, P. T., \& Turner, M. J. L., 2003, MNRAS, 345, 705

\bibitem[Proga \& Kallman(2004)]{PK04} 
Proga, D., \& Kallman, T. R., 2004, ApJ, 616, 688

\bibitem[Proga et al.(2000)]{Proga00}
Proga, D., Stone, J. M., \& Kallman, T. R. 2000, ApJ, 543, 686

\bibitem[Reeves et al.(2008)]{Reeves08}
Reeves, J. N., Done, C., Pounds, K. A., Terashima, Y., Hayashida, K., 
Anabuki, N., Uchino, M., \& Turner, M. J. L., 2008, MNRAS, 385, L108

\bibitem[Reeves et al.(2003)]{Reeves03} 
Reeves, J. N., O'Brien, P. T., \& Ward, M. J., 2003, ApJ, 593, L65

\bibitem[Reeves et al.(2002)]{Reeves02}
Reeves, J. N., Wynn, G., O'Brien, P. T., \& Pounds, K. A., 2002, MNRAS, 336, L56

\bibitem[Reeves et al.(2000)]{Reeves00} 
Reeves, J. N., O'Brien, P. T., Vaughan, S., Law-Green, D.,
Ward, M., Simpson, C., Pounds, K. A., \& Edelson, R. 2000, MNRAS, 312, L17

\bibitem[Revnivtsev et al.(2006)]{Rev06}
Revnivtsev, M., Sazonov, S., Gilfanov, M., Churazov, E., \& Sunyaev, R., 
2006, A\&A, 452, 169 

\bibitem[Revnivtsev et al.(2004)]{Rev04}
Revnivtsev, M., Sazonov, S., Jahoda, K., \& Gilfanov, M., 2004, A\&A, 418, 927 

\bibitem[Risaliti(2002)]{Risaliti02}
Risaliti, G. 2002, A\&A, 386, 379

\bibitem[Ross \& Fabian(2005)]{RF05}
Ross, R. R., \& Fabian, A. C. 2005, MNRAS, 358, 211

\bibitem[Ross et al.(1999)]{Ross99}
Ross, R. R., Fabian, A. C., \& Young, A. J., 1999, MNRAS, 306, 461

\bibitem[Schurch et al.(2009)]{Schurch09}
Schurch, N. J., Done, C., \& Proga, D., 2009, ApJ, 694, 1

\bibitem[Silk \& Rees(1998)]{SR98}
Silk, J., \& Rees, M. J., 1998, A\&A, 331, L1

\bibitem[Sim et al.(2008)]{Sim08}
Sim, S. A., Long, K.S., Miller, L., \& Turner, T.J., 2008, MNRAS, 388, 611

\bibitem[Simpson et al.(1999)]{Simpson99} 
Simpson, C., Ward, M., O'Brien, P. T., \& Reeves, J. N. 1999,
MNRAS, 303, L23

\bibitem[Takahashi et al.(2007)]{Takahashi07}
Takahashi, T., et al., 2007, PASJ, 59, 35

\bibitem[Torres et al.(1997)]{Torres97} 
Torres, C. A. O., et al., 1997, ApJ, 488, L19

\bibitem[Tueller et al.(2008)]{Tueller08}
Tueller, J., Mushotzky, R. F., Barthelmy, S., Cannizzo, J. K., Gehrels, N., Markwardt, 
C. B., Skinner, G. K., \& Winter, L. M., 2008, ApJ, 681, 113

\bibitem[Turner et al.(2009)]{Turner09} 
Turner, T. J., Miller, L., Kraemer, S. B., Reeves, J. N., 
\& Pounds, K. A., 2009, ApJ, submitted

\bibitem[Vignali et al.(2000)]{Vignali2000}
Vignali, C., Comastri, A., Nicastro, F., Matt, G., Fiore, F., \& Palumbo, G.G.C., 
2000, A\&A, 362, 69

\bibitem[Wilms et al.(2000)]{Wilms00}
Wilms, J., Allen, A., \& McCray, R. 2000, ApJ, 542, 914

\end{thebibliography}
\end{document}